\documentclass[12pt,aps,tightenlines,notitlepage,superscriptaddress]{revtex4-1}

\usepackage{lineno,hyperref}
\modulolinenumbers[5]
\usepackage{amsmath}
\usepackage{amssymb}
\usepackage{bm}
\usepackage[mathscr]{eucal}
\usepackage{theorem}
\usepackage{graphicx}
\usepackage{psfrag}
\usepackage{subfigure}
\usepackage{color}
\usepackage[table]{xcolor}
\usepackage{wasysym}
\include{graphix}
\usepackage{framed}
\usepackage{enumitem}%
\usepackage{lipsum} 
\usepackage{float}
\usepackage{stmaryrd}
\usepackage[a4paper]{geometry}
\usepackage{float}
\usepackage{ulem}

\graphicspath{{figs/}}

\newcommand{\mathd}{\mathrm{d}}

\newcommand{\vecx}{\bm{x}}
\newcommand{\vecu}{\bm{u}}

\newcommand{\vecg}{\bm{g}}

\newcommand{\vecsig}{\bm{\sigma}}

\newcommand{\vecI}{\bm{I}}

\newcommand{\rf}{\rho_{\mathrm{f}}}
\newcommand{\rp}{\rho_{\mathrm{p}}}

\newcommand{\dc}{D_{\mathrm{c}}}
\newcommand{\dv}{D_{\mathrm{v}}}

\newcommand{\rRe}{\mathrm{Re}}

\newcommand{\rFr}{\mathrm{Fr}}
\newcommand{\rBn}{\mathrm{Bn}}

\DeclareMathAlphabet{\mathpzc}{OT1}{pzc}{m}{it}

\newcommand{\mygrav}{\mathcal{G}}
\newcommand{\mystress}{\bm{T}}

\begin{document}

\title{Particle-laden viscous channel flows -- model regularization and parameter study}

\author{Lennon \'O N\'araigh}
\email{onaraigh@maths.ucd.ie}

\affiliation{School of Mathematics and Statistics, University College Dublin, Belfield, Dublin 4}
\affiliation{Complex and Adaptive Systems Laboratory, University College Dublin, Belfield, Dublin 4}

\author{Ricardo Barros}

\affiliation{Mathematics Applications Consortium for Science and Industry (MACSI), Department of Mathematics and Statistics, University of Limerick, Limerick, Ireland}

\date{\today}

\begin{abstract}
We characterize the flow of a viscous suspension in an inclined channel where the flow is maintained in a steady state under the competing influences of gravity and an applied pressure drop.  The basic model relies on a diffusive-flux formalism.  Such models are common in the literature, yet many of them possess an unphysical singularity at the channel centreline where the shear rate vanishes.  We therefore present a regularization of the basic diffusive-flux model that removes this singularity.  This introduces an explicit (physical) dependence on the particle size into the model equations.  This approach enables us to carry out a detailed parameter study showing in particular the opposing effects of the pressure drop and  gravity.  Conditions for counter-current flow and complete flow reversal are obtained from numerical solutions of the model equations.  These are supplemented by an analytic lower bound on the ratio of the gravitational force to the applied pressure drop necessary to bring about complete flow reversal.   
\end{abstract}


\maketitle

\section{Introduction}
\label{sec:intro}

Particles suspended in viscous flow occur in a wide variety of applications.  In certain technical operations (for example drilling oil wells), it is important to be able to predict the properties of the suspension as a function of flow rate, particle size, etc., with a view to controlling the operation in real time~\cite{larsen1990}.  There is therefore a strong motivation to develop accurate models to characterize the hydrodynamics of the suspension.  In this work we introduce a simple model to characterize the flow of suspension in an inclined channel under equilibrium conditions.  The modelling framework is the diffusive-flux model.  Such models certainly abound in the literature -- and many of them exhibit an unphysical singularity in the shear rate at the channel centreline.  The main goal of this work therefore is to introduce a self-consistent regularization that removes this singularity.  A second goal is to carry out a detailed parameter study based on the regularized model to fully characterize the hydrodynamics as a function of the dimensionless parameters in the problem -- both in the horizontal and inclined cases.  Before doing this, we place our work in the context of the existing literature on the subject.

There are at least two distinct approaches to modelling a suspension of dense particles in a (Newtonian) liquid.   In the first approach, called the suspension balance model, the averaged dynamics of the suspended particles are described in a statistical-mechanics formalism.   However, the model couples to the fluid mechanics of the problem in a natural way.   This model was first proposed in Reference~\cite{nott1994}.  A review of the model (along with various refinements thereto) can be found in References~\cite{morris1998,fang2002}.  The model involves mass and momentum equations for the particle phase (averaged over a test volume) and the mixture (again averaged over a test volume), leading to four evolutionary equations in the first instance.  Both momentum equations involve particle-phase and mixture stress tensors respectively, and the particle momentum equation further involves a hydrodynamic drag force, meaning that three constitutive relations are required for closure.  The closure is achieved by modelling the hydrodynamic drag force and various viscous terms.  The particle-phase shear stress term is modelled by the introduction of an auxiliary variable (the particle-phase `temperature'), leading to a set of five coupled evolution equations.
A simpler approach that makes predictions of comparable accuracy to the suspension-balance model is the diffusive-flux model, first introduced in Reference~\cite{phillips1992} but based partly on earlier work~\cite{leighton1987} (see also Reference~\cite{schaflinger1993}).  
The idea here is to focus entirely on the mixture for the hydrodynamic model, together with an  advection-diffusion equation for the volume fraction $\phi$ of the particles. The particle flux in this equation is then modelled according to the collision dynamics of the particles, to include shear-induced migration, viscous migration, and gravitational settling.

In the present work, the diffusive-flux model is adopted.    The reasons for this choice are manifold: the diffusive flux framework is both conceptually and analytically straightforward, and involves only a handful of parameters, all of which can be estimated from benchmark cases.  Although it has shortcomings~\cite{fang2002}, it produces acceptable results for flow profiles and volume profiles in horizontal pressure-driven pipe/channel flows, as well as in rotating shear flow~\cite{phillips1992}.  Finally, it has been shown that the suspension balance and diffusive-flux models share the same basic framework, the main difference being the choice of closure relations for the different parameters~\cite{vollebregt2010}.  
%
%
%

In spite of the tractability of the diffusive-flux model, in its basic form it cannot be applied to fully-developed flow in an inclined channel.  This is because the model develops a singularity wherever the shear rate vanishes.  A review of the literature shows that this problem is overcome in certain highly specific contexts -- e.g. resorting to a symmetry and placing the singularity at the centreline of a horizontal pipe/channel~\cite{phillips1992}, or exploiting the specific properties of interfacial flows and
placing the singularity at a free surface~\cite{murisic2011}.  Yet the geometry of an inclined channel prevents these solutions from being applied in the present context.
Furthermore, existing efforts to overcome these issues are incomplete.  Reference~\cite{schaflinger1993} looks at inclined flows, but only in the context of Brownian particle diffusion, which is not relevant at the high P\'eclet numbers with which this work is concerned and in any case is not a diffusive-flux model.  Reference~\cite{kauzlaric2011non} introduces a regularization of the full diffusive-flux model that removes the singularity  through the introduction of a collision rate proportional to a shear rate that is averaged over a particle radius.  However, the averaging is accomplished using effectively an $L^1$ norm, which on mathematical grounds is not optimal, as such an approach does not completely regularize the model.  More importantly, the regularized model is not applied to inclined flows.
%
%
%
%
%
%
%
%
Therefore, a main aim of the present work  is to derive a regularization procedure that fully heals the singularity inherent in diffusive-flux models.  This then enables a full parameter study for inclined flows that takes account of the different flow regimes that arise as a result of the competing effects of the pressure drop and gravity, as well as the bulk volume fraction and the channel inclination.  

This work is organized as follows.  In Section~\ref{sec:genframe} the standard diffusive-flux model from the literature is summarized.  A diffusive-flux model specific to steady-state operations  in inclined channel flow is presented in Section~\ref{sec:mathmod}, along with a regularization to heal the singularity that would otherwise occur where the shear-rate vanishes.  Results based on this approach are presented in Section~\ref{sec:results}, including a detailed parameter study outlining the conditions under which different flow regimes are observed.  We discuss the application of our model to suspending fluids with non-Newtonian rheology in Section~\ref{sec:disc}, wherein concluding remarks are also given.


\section{General theoretical framework}
\label{sec:genframe}

In this section we summarize the full diffusive-flux theoretical  framework existant in the literature, with  a view later on to subject this model to a regularization technique to enable a full parameter study of the steady-state flow in an inclined channel.
The starting point is a momentum equation for the velocity $\vecu(\vecx,t)$  of a parcel comprising a mixture of particles and suspending fluid:
\begin{equation}
\rho(\phi)\left(\frac{\partial\vecu}{\partial t}+\vecu\cdot\nabla\vecu\right)=\nabla\cdot\mystress+\rho(\phi)\vecg.
\label{eq:hydro}
\end{equation}
where $\phi$ is the particle-phase volume fraction,  $\vecg$ is the acceleration due to gravity and $\mystress$ is the mixture stress tensor.
Furthermore, the density is given by 
\begin{equation}
\rho(\phi)=\rp\phi+\rf(1-\phi),
\end{equation}
whre
$\rf$ is the constant fluid density and $\rp$ is the particle density, also constant.
This is supplemented by the incompressibility condition $\nabla\cdot\vecu=0$.  
 The evolution of the volume fraction $\phi$ is given by a flux-conservative equation,
\begin{equation}
\frac{\partial\phi}{\partial t}+\vecu\cdot\nabla\phi=-\nabla\cdot\bm{J}_\phi,
\label{eq:flux}
\end{equation}
where the particle flux $\bm{J}_\phi$ is modelled according to the collision dynamics of the particles, to include shear-induced migration, viscous migration, and gravitational settling.  

A classification of the collective particle dynamics provide a means of constituting the flux $\bm{J}_\phi$.  The main effect to consider is shear-induced migration, which is based on the observation that in a  dense suspension, particles that are transported by a shear flow will collide.   The collision rate is proportional to $\phi\dot\gamma$, where $\dot\gamma$ is the (unsigned) local shear rate.  Particles will move from regions where the collision rate is high to a nearby region where the collision rate is lower, meaning that there is a shear-induced contribution $\bm{J}_\mathrm{c}$ to the total flux, with $\bm{J}_\mathrm{c}\propto -\nabla (\phi\dot\gamma)$.  Reference~\cite{phillips1992} gives the shear-induced flux as
\begin{equation}
\bm{J}_\mathrm{c}=-\dc \phi a^2\nabla\left(\phi\dot\gamma\right),
\end{equation}
where $\dc$ is a dimensionless constant and $a$ is the particle radius (a monodisperse suspension of identical spherical particles is assumed).  A second effect is present in viscous flows, whereby particles will move into regions of lower viscosity, from regions of higher viscosity.  In Reference~\cite{phillips1992} this is modelled in such a way that that the viscous contribution to the total flux is proportional to the  ratio between the viscosity gradient (giving the direction of migration) and the local viscosity, giving a total contribution
\begin{equation}
\bm{J}_\mathrm{v}=-\dv a^2\phi^2\dot\gamma\left(\frac{\nabla\mu}{\mu}\right),
\end{equation}
where $\dv$ is another dimensionless constant.

For particles whose density is greater than that of the suspending fluid, settling will occur, leading to a gravitational flux.  For Stokes flow, this can be modelled as
\begin{equation}
\bm{J}_\mathrm{g}=\frac{2a^2(\rp-\rf)\phi f(\phi)}{9\mu_\mathrm{f}}\bm{g},
\label{eq:j_grav}
\end{equation}
where $\mu_\mathrm{f}$ is the (constant) dynamic viscosity of the suspending fluid, and $f(\phi)$ is the so-called `hindrance function', introduced here because the collective settling flux in a suspension differs from the corresponding single-particle expression because neighbouring particles `hinder' a given particle's descent through the medium.  Using the same reasoning, walls also hinder settling, and exact expressions for single-particle settling in the neighbourhood of a wall are known~\cite{happel1983}; these effects may be parametrized through a modification of Equation~\eqref{eq:j_grav}:
\begin{equation}
\bm{J}_\mathrm{g}=\frac{2a^2(\rp-\rf)\phi f(\phi)}{9\mu_\mathrm{f}}\bm{g}\,\omega(z).
\label{eq:j_grav1}
\end{equation}
Expressions for $\omega(z)$ can be found in the literature. Here we use 
$$\omega(z)=A(z/a)^2 / \sqrt{1+A^2 (z/a)^4},$$ 
with $A=1/6$, (Reference~\cite{murisic2011} uses $A=1/18$, but we have verified that the results contained herein are insenstive to this choice), so that $\omega(z)\rightarrow 0$ as $z\rightarrow 0$ and $\omega\approx 1$ away from $z=0$. Also, numerous (and very similar) forms exist for the hindrance function (see e.g. Table 3 in Reference~\cite{vollebregt2010}); here we use $f(\phi)=(1-\phi)\mu_\mathrm{f}/\mu(\phi)$, where $\mu(\phi)$ is the effective viscosity of the suspension.  For dense suspensions, the Krieger--Dougherty relation is appropriate here, giving the effective suspension viscosity as
\begin{equation}
\mu(\phi)=\mu_\mathrm{f}\left(1-\frac{\phi}{\phi_\mathrm{m}}\right)^{-\xi},
\end{equation}
where $\xi$ is a positive constant and $\phi_\mathrm{m}>0$ is the maximum volume fraction achievable by the spherical particles.  Following standard practice~\cite{murisic2011}, we take $\xi=2$ in this work.

For completeness, it is noted that the particles in the suspension will experience thermal fluctuations, giving rise to a purely Brownian flux term $\bm{J}_\mathrm{d}=-D\nabla \phi$, where $D$ is the diffusivity.  The importance of the diffusivity is estimated through the particle P\'eclet number, $\mathrm{Pe}=\dot\gamma a^2/D$.  For the present applications, this is typically a large number~\cite{phillips1992}, meaning that the Brownian contribution to the total flux can be ignored.
In this way, the particle flux $\bm{J}_\phi$ is modelled as 
\begin{equation}
\bm{J}_\phi=\bm{J}_{\mathrm{c}} + \bm{J}_{\mathrm{v}} + \bm{J}_{\mathrm{g}}+\bm{J}_\mathrm{d},
\label{eq:jphi}
\end{equation}
with the final (Brownian) contribution neglected in what follows.

\section{Specific mathematical model including model regularization}
\label{sec:mathmod}

\begin{figure}[htb]
  \begin{center}
    \includegraphics[width=0.8\textwidth]{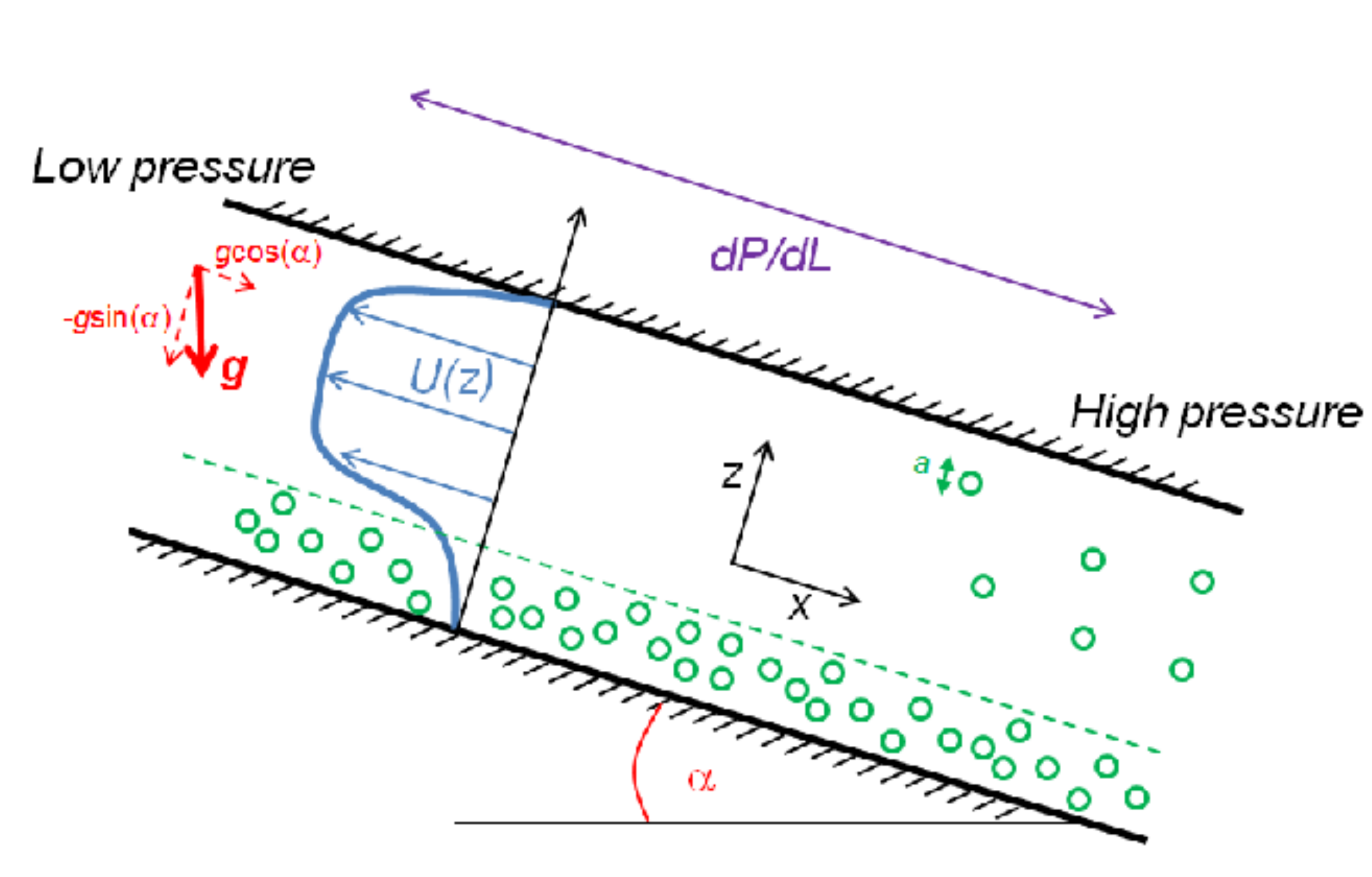}
  \end{center}
  \caption{Schematic description of the model problem}
	\label{fig:sketch_channel}
\end{figure}
We consider pressure-gravity-driven flow in an inclined channel under equilibrium conditions as shown in Figure~\ref{fig:sketch_channel}.
The focus of the present work is on the equilibrium scenario wherein the mixture shear stress balances with the pressure drop and the gravitational force.  The reasons for this are manifold:
 it is a simple scenario amenable to a semi-analytical description; it is a scenario wherein diffusive-flux models are known to produce acceptable results for the hitherto-investigated horizontal case, and finally, it is an important base case that can be generalized in the future to include a fully transient flow.


We start by assuming that the total mixture stress tensor is given by $\mystress=-p \vecI + \vecsig$,
in which $p$ is the pressure, and $\vecsig=\mu(\phi)\dot{\bm{\gamma}}$.  Here $\dot{\bm{\gamma}}$ is the rate-of-strain tensor with components $\dot{\gamma}_{ij}$, where
\begin{equation}
\dot{\gamma}_{ij}=\frac{\partial u_i}{\partial x_j}+\frac{\partial u_j}{\partial x_i}.
\label{eq:sigmagamma}
\end{equation}
%
%
%
The corresponding unsigned local rate of strain is given by
\begin{equation}
\dot\gamma=\sqrt{\dot\gamma_{ij}\dot\gamma_{ij}},
\label{eq:gamma_unsigned}
\end{equation}
where we sum over repeated indices.  
%
%

Under these assumptions, a fully developed flow corresponds to an equilibrium situation wherein the pressure drop balances with the tangential stress and gravity force, such that
the relevant diffusive-flux equations read
\begin{subequations}
\begin{eqnarray}
\frac{\mathd\sigma}{\mathd z}&=&\frac{dP}{dL}-\rho(\phi) g\sin\alpha,\\
0&=&J_\mathrm{c}+J_\mathrm{v}+J_\mathrm{g},
\end{eqnarray}
\end{subequations}
where $\sigma=\pm\mu(\phi)\dot\gamma$ is the (signed) shear stress of the mixture, and $\dot\gamma=|\mathd U/\mathd z|$ is the rate of strain.  Here, the suspending fluid is assumed to be Newtonian; the effects of a non-Newtonian rheology in the suspending fluid are discussed briefly in Section~\ref{sec:disc}.
We nondimensionalize the equations of motion based on the channel height $H$ and the characteristic velocity $V$, where 
\[
V=\frac{H^2}{\mu_\mathrm{f}}\frac{dP}{dL}, 
\]
thereby introducing dimensionless variables $\tilde z=z/H$, $\tilde U=U/V$, $\widetilde{\dot\gamma}=\dot\gamma(H/V)$ and $\tilde\sigma=\sigma/\sigma_0$, with $\sigma_0=\mu_\mathrm{f}V/H$.  Also, all densities are scaled relative to the density $\rf$ of the suspending fluid. Based on these scaling rules, and based on the closure relations for the fluxes described in Section~\ref{sec:genframe}, the following non-dimensional equations of motion are obtained:
%
%
%
\begin{subequations}
\begin{eqnarray}
\frac{\mathd\tilde\sigma}{\mathd \tilde z}&=&1-\tilde\rho(\phi) \rRe \, \rFr^{-2}\sin\alpha,\label{eq:sigma_fully_d}\\
0&=&\dc  \phi\frac{\mathd}{\mathd \tilde z}\left(\widetilde{\dot\gamma}\phi\right)+
\dv \phi^2\widetilde{\dot\gamma} \frac{1}{\mu}\frac{\mathd\mu}{\mathd \tilde z}\nonumber\\
&\phantom{=}&\phantom{aaaaaaaaaaaaaaaaa}+\frac{2(r-1)\phi(1-\phi)}{9\tilde\mu(\phi)}\rRe \, \rFr^{-2}\cos\alpha\,\omega(\tilde{z}),
\label{eq:J_fully_d}
\end{eqnarray}
\end{subequations}
where $r=\rp/\rf$, $\tilde{\rho}=r\phi+(1-\phi)$, $\tilde\mu(\phi)=[1-(\phi/\phi_\mathrm{m})]^{-\xi}$,  and where
\[
\rRe=\frac{V H \rf}{\mu_\mathrm{f}},\qquad \rFr^{-2}=\frac{ g H}{V^2}.
\]
Following standard practice, the ornamentation over the dimensionless variables is now dropped.  We next rewrite Equation~\eqref{eq:J_fully_d} in terms of $\sigma$, removing all instances of $\dot\gamma$ as follows:
\begin{equation}
\dc  \phi\frac{\mathd}{\mathd z}\left(\frac{|\sigma|}{\mu}\phi\right)+\dv \phi \frac{|\sigma|}{\mu^2}\frac{\mathd\mu}{\mathd z}+
\frac{2(r-1)\phi(1-\phi)}{9\mu(\phi)}\rRe \,\rFr^{-2}\cos\alpha\,\omega(z)=0.
\label{eq:J_full_d_gamma}
\end{equation}
A simple rearrangement of terms, 
by using the Krieger--Dougherty relation $\mu=[1-(\phi/\phi_{\mathrm{m}})]^{-2}$ and computing $\mu$-derivatives,
leads to 
\begin{equation}
\frac{\mathd\phi}{\mathd z}=\frac{-\phi \frac{\mathd|\sigma|}{\mathd z}-\frac{2}{9\dc}\rRe \,\rFr^{-2}(r-1)(1-\phi)\cos\alpha\,\omega(z)}{|\sigma|\left[1+2\left(\frac{\dv-\dc}{\dc}\right)\frac{\phi}{\phi_\mathrm{m}-\phi}\right]}.
\label{eq:dphidz_dodgy}
\end{equation}
Note that all \textit{explicit} dependence of the problem on the particle size  has dropped out, as neither Equation~\eqref{eq:J_fully_d} nor Equation~\eqref{eq:dphidz_dodgy} exhibits an explicit dependence on $a$ (there is however some implicit dependence, via the chosen functional form of $\omega(z)$).  This issue is commonly encountered in diffusive-flux-type models, yet is unphysical.  This problem, as well as others described below,  mean that it is necessary to regularize Equation~\eqref{eq:dphidz_dodgy}, which is the subject of the remainder of this section.


In practice, it is ill-advised to attempt to solve Equation~\eqref{eq:dphidz_dodgy} because of the singularity at places where the shear stress vanishes (corresponding to the centreline in single-phase Poiseuille flow). This could be fixed by letting the collision rate go to zero when the shear rate vanishes; however, this is unphysical. 
An alternative would be proposing that ``non-locality'' is required in order to capture the collision of particles with a deformation rate which is locally vanishing. To do so, we consider 
the (unsigned) shear stress averaged over a single spherical particle of radius $a$ (see \ref{sec:app:derivation}):
\begin{equation}
\hat\sigma=\sqrt{\sigma^2+\tfrac{1}{3}a^2\left(\frac{\mathd\sigma}{\mathd z}\right)^2}.
\label{eq:sig_avgxxx}
\end{equation}
%

Equation~\eqref{eq:sig_avgxxx} and its implementation in the model equation~\eqref{eq:dphidz_dodgy} can be regarded as a non-local improvement of the basic Phillips model for an inclined flow.  Effectively, Equation~\eqref{eq:sig_avgxxx} is the root-mean-square average shear stress over a single particle.
Use of the $L^2$ norm (and hence the root-mean-square average) in Equation~\eqref{eq:sig_avgxxx} is justified   because it renders the following calculations -- in particular, integrals -- straightforward.  More importantly, this approach means that the  shear stress appears in a differentiable fashion  in the final equation set.  In this way, all non-differentiable contributions to the model equations (e.g. $1/|\sigma|$ and $\mathd|\sigma|/\mathd z$ in Equation~\eqref{eq:dphidz_dodgy}) are regularized.
%
%
%
Crucially, this can be readily extended to out-of-equilibrium scenarios well beyond the simple scenario illustrated in Figure~\ref{fig:sketch_channel}.

To understand how Equation~\eqref{eq:sig_avgxxx} is worked into a balance model for the shear stress and the volume fraction profiles, we start with the unregularized Equation~\eqref{eq:J_full_d_gamma}
and replace all instances of $|\sigma|$  with $\hat\sigma$, to yield
\begin{equation}
\dc  \phi\frac{\mathd}{\mathd \tilde z}\left(\frac{\hat\sigma}{\mu}\phi\right)+\dv \phi^2 \frac{\hat\sigma}{\mu^2}\frac{\mathd\mu}{\mathd z}+
\frac{2(r-1)\phi(1-\phi)}{9\mu(\phi)}\rRe \, \rFr^{-2}\cos\alpha\,\omega(z)=0.
\label{eq:J_full_d_gamma1}
\end{equation}
It is worth to observe that the 
Reynolds and Froude numbers combine together as
\begin{equation}
\rRe\,\rFr^{-2}=\frac{\rho g}{|\mathd P/\mathd L|}.
\label{eq:combo}
\end{equation}
Equivalently, one may introduce the velocity $\hat{V}=\sqrt{(H/\rho)|\mathd P/\mathd L|}$, such that
$\rRe \,\rFr^{-2}= gH/{\hat{V}^2}$ and the dimensionless group can be identified as the inverse of a squared Froude number (based on the rescaled velocity $\hat{V}$).
%
%
Throughout the remainder of the work, we therefore use the dimensionless quantity
\begin{equation}
\mygrav:=\rRe\,\rFr^{-2}
\label{eq:gdef}
\end{equation}
as the key parameter. Notice that $\mygrav (r-1)$ is nothing more than a local Richardson number.


We proceed with calculations and reduce Equation~\eqref{eq:J_full_d_gamma1} to
\begin{equation}
\left[1+\frac{2(\dv-\dc)}{\dc}\frac{\phi}{\phi_\mathrm{m}-\phi}\right]\hat\sigma\frac{\mathd\phi}{\mathd z}=-\frac{\mathd\hat\sigma}{\mathd z}\phi-\tfrac{2}{9\dc}\mygrav(r-1)\cos\alpha(1-\phi)\omega(z),
\label{eq:J_full_d_gamma2}
\end{equation}
%
%
where the dimensionless average shear stress identified as 
\begin{equation}
\hat\sigma=\sqrt{\sigma^2+\epsilon^2\left(\frac{\mathd\sigma}{\mathd z}\right)^2},\qquad \epsilon=\tfrac{1}{\sqrt{3}}(a/H).
\label{eq:sig_avgxxx_nondim}
\end{equation}
%
%
Using Equations~\eqref{eq:sigma_fully_d} and~\eqref{eq:sig_avgxxx_nondim}, we obtain
\begin{eqnarray}
\frac{\mathd\hat\sigma}{\mathd z}&=&\frac{1}{\hat\sigma}\left(\sigma \frac{\mathd\sigma}{\mathd z}+\epsilon^2\frac{\mathd\sigma}{\mathd z}\frac{\mathd^2\sigma}{\mathd z^2}\right),\nonumber\\
&=& \frac{1}{\hat\sigma} \frac{\mathd\sigma}{\mathd z} \left( \sigma + \epsilon^2 \frac{\mathd^2\sigma}{\mathd z^2} \right), \nonumber\\
%
&=&\frac{\sigma}{\hat\sigma}\frac{\mathd\sigma}{\mathd z}-\frac{\epsilon^2}{\hat\sigma}\frac{\mathd\sigma}{\mathd z}\left[\mygrav\sin\alpha(r-1)\frac{\mathd\phi}{\mathd z}\right],\label{eq:dsigmahat}
\end{eqnarray}
where $\sigma$ is the signed shear stress.
%
Further regularization  of the $\phi$-equation is applied whenever $\phi=0$ or $\phi=\phi_\mathrm{m}$: in those cases, $\mathd \phi/\mathd z$ is set to zero.  This forces $\phi$ to remain within the physical bounds $0\leq \phi \leq \phi_\mathrm{m}$.  Thus, the fully consistent regularized equation set reads
\begin{subequations}
\begin{eqnarray}
\frac{\mathd U}{\mathd z}&=&\frac{\sigma}{\mu},\label{eq:dudz}\\
\frac{\mathd\sigma}{\mathd z}&=&1-\mygrav\left[r\phi+(1-\phi)\right]\sin\alpha,\label{eq:dsigdz}\\
\frac{\mathd\phi}{\mathd z}&=&
\begin{cases}0,&\text{if }\phi=0,\text{ or }\phi=\phi_\mathrm{m},\\
\frac{-\phi \frac{\sigma}{\hat\sigma}\frac{\mathd\sigma}{\mathd z}-\frac{2}{9\dc} \mygrav(r-1)(1-\phi)\cos\alpha\,\omega(z)}{\hat\sigma\left[1+2\left(\frac{\dv-\dc}{\dc}\right)\frac{\phi}{\phi_\mathrm{m}-\phi}-\frac{\epsilon^2}{\hat\sigma^2}\frac{\mathd\sigma}{\mathd z} \phi \,\mygrav(r-1)\sin\alpha\right]},&\text{otherwise}\end{cases}
\end{eqnarray}%
\label{eq:model_ode}%
\end{subequations}%

\section{Results and parameter study}
\label{sec:results}

 Equation~\eqref{eq:model_ode} is a two-point boundary value problem involving three first-order ordinary differential equations.  Two boundary conditions are obvious: $U(0)=U(1)=0$, corresponding to no-slip at the channel walls.  In practice, the third boundary condition is prescribed as $\phi(0)=\phi_1$, and $\phi_1$ is adjusted until the  corresponding prescribed bulk cuttings volume fraction $\Phi$ is obtained, where
\begin{equation}
\Phi=\int_0^1 \phi(z)\,\mathd z.
\end{equation}
In the present section, we report on results wherein the  model ordinary differential equations (ODEs)~\eqref{eq:model_ode} are solved numerically using a shooting method.  Boundary conditions at $(U(0)=0,\sigma(0)=\sigma_1,\phi(0)=\phi_1)$ are supplied and the parameter $\sigma_1$ is adjusted using a rootfinding procedure until the no-slip condition at $z=1$ is also satisfied.    Also, the functional form for $\omega(z)$ is taken directly from Reference~\cite{murisic2011}, which, when expressed in terms of dimensionless quantities, reads as 
$$
\omega(z) = A z^2 / \sqrt{9 \epsilon^4 + A^2 z^4}.
$$
A full characterization of all the solutions to Equations~\eqref{eq:model_ode} requires the exploration of a multidimensional parameter space involving the five independent parameters $(\alpha,\epsilon,\Phi,\mygrav,r)$, with $\dc=0.43$, $\dv=0.65$, $\xi=2$, $\phi_\mathrm{m}=0.68$ set by theory.  Throughout this section, we set $\epsilon=0.01$.  We also initially set $\alpha=\pi/12$ and $r=2$ and 
 focus in the first instance on the parameter subspace $(\Phi,\mygrav)$.  However, we also subsequently investigate the effects of varying angle of inclination and density ratio -- see Section~\ref{sec:flow_pattern_map} below.

\subsection{Sample results}

\begin{figure}[H]
	\centering
		\subfigure[$\,\,\mygrav=0.1$]{\includegraphics[width=0.32\textwidth]{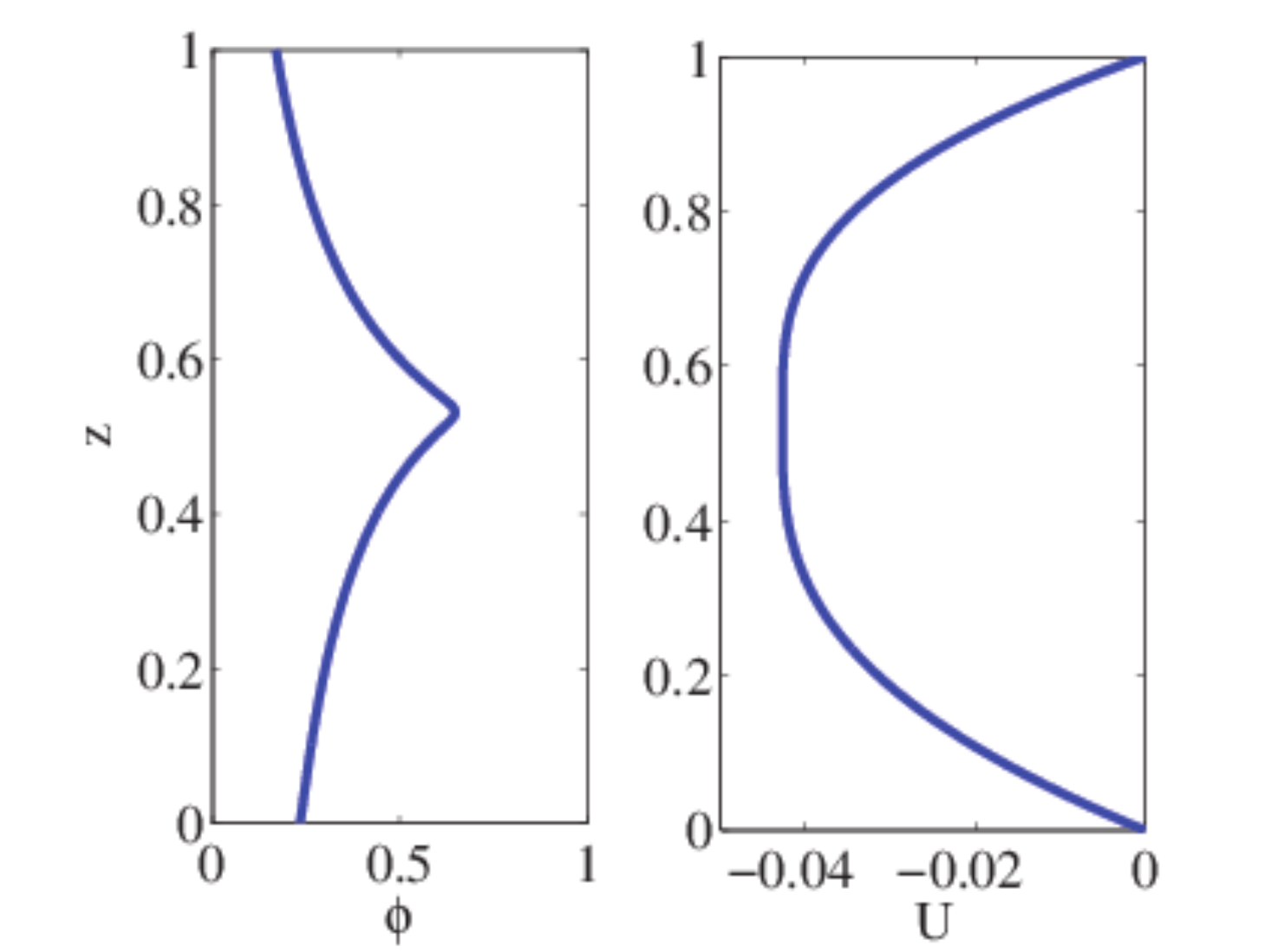}}
		\subfigure[$\,\,\mygrav=2$]{\includegraphics[width=0.32\textwidth]{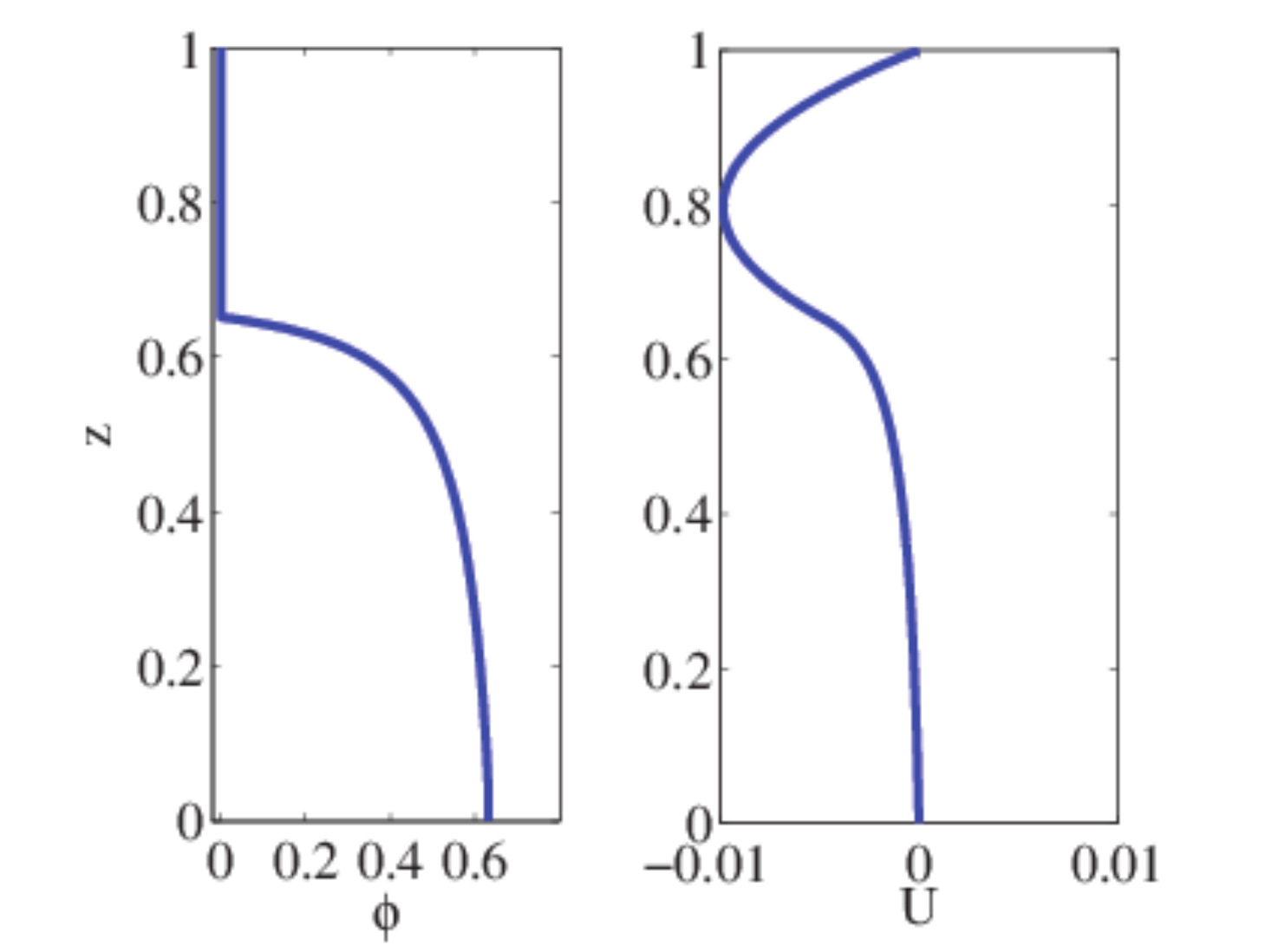}}
		\subfigure[$\,\,\mygrav=10$]{\includegraphics[width=0.32\textwidth]{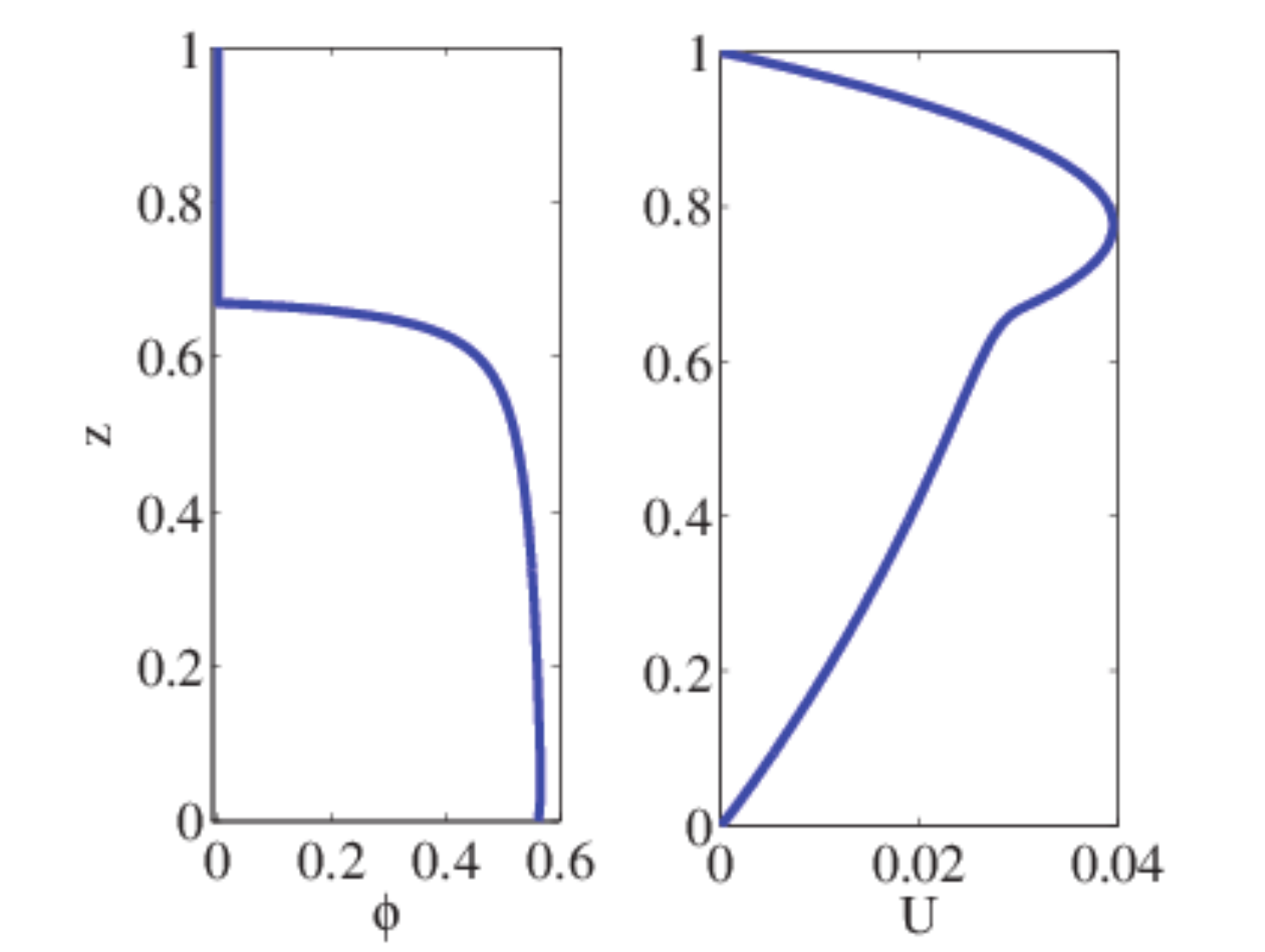}}
		\caption{Sample profiles for $\Phi=0.35$ and various values of $\mygrav$, corresponding to (a) weak gravity effect, (b) gravity effect finely balanced compared to pressure effect and (c) strong gravity effect.  Increasing the gravity effect leads to flow reversal, corresponding to a change in the direction of the flow profile in (c). The values for the physical parameters $r$, $\epsilon$, and $\alpha$ will be fixed throughout the text as $r=2$, $\epsilon=0.01$, and $\alpha=\pi/12$.}
	\label{fig:sample_figs}
\end{figure}
Sample results are shown in Figure~\ref{fig:sample_figs} for the case $\Phi=0.35$ and various values of $\mygrav$.  Here, in panel (a), the effect of gravity is small compared to the applied pressure gradient, and the mixture flows up the channel under the applied pressure gradient.  The volume-fraction profile and the mixture flow profile are similar to those observed in pure pressure-driven channel flows, and the system is `well mixed', in the sense that a nonzero volume fraction extends from the bottom wall to the top wall, with a maximum distribution of particles close to the channel centreline, corresponding to shear-induced migration.  The volume-fraction profile possesses a single sharp maximum: close inspection shows that this is not a `kink' and that the profile is smooth over lengthscales comparable to $\epsilon$.  Increasing $\epsilon$ (not shown) causes the profile to smoothen further over a wider range of scales, confirming the efficacy of the regularization introduced in Equation~\eqref{eq:model_ode}.

\begin{figure}
	\centering
		\includegraphics[width=0.5\textwidth]{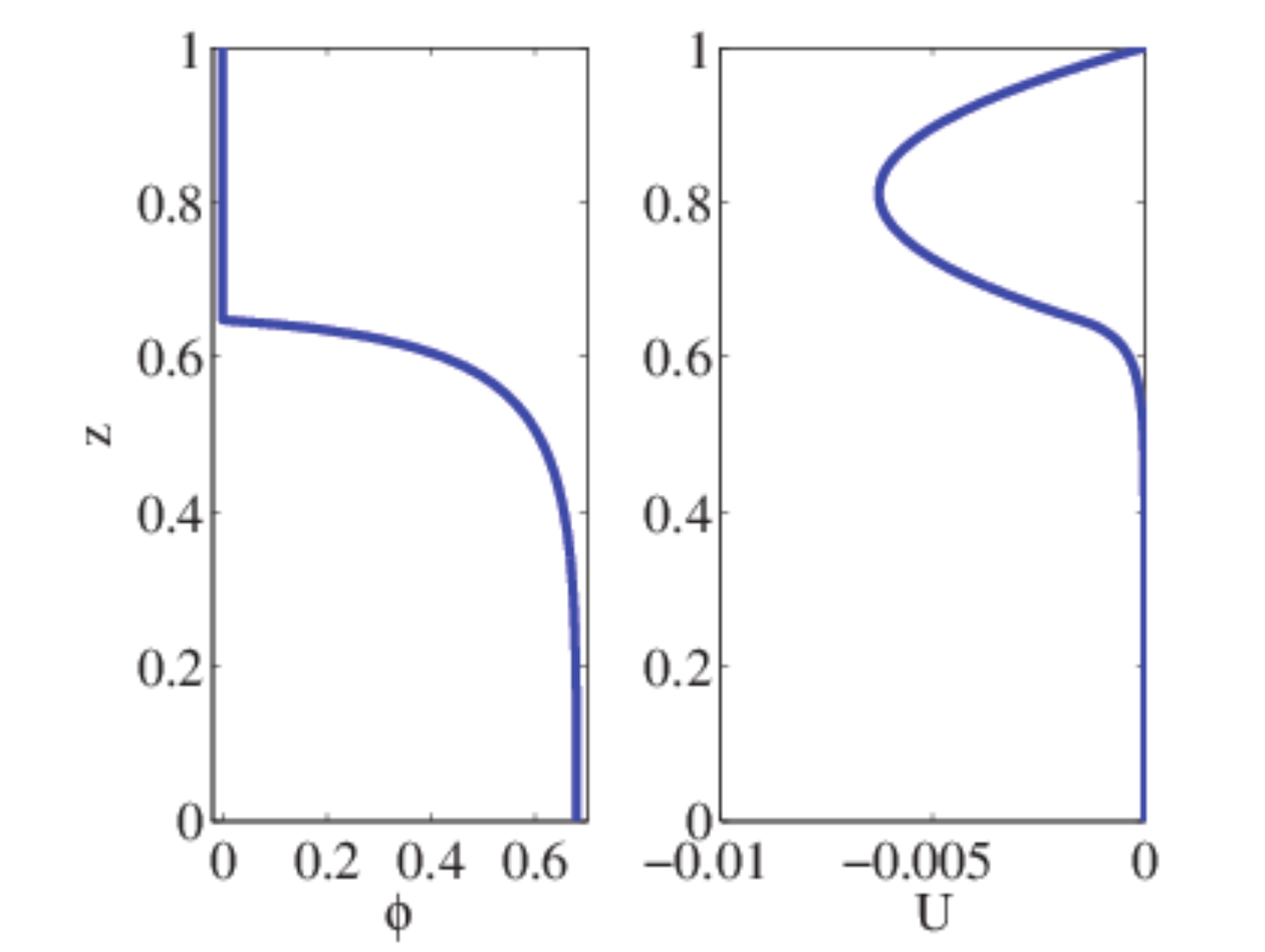}
		\caption{Stationary particle bed with clear layer of fluid transported upward ($\Phi=0.4, \mygrav=2.5$).}
	\label{fig:sample_figs1}
\end{figure}

Upon increasing the gravity effect compared to the pressure effect (panel (b)), the system ceases to be well mixed, the particles settle, and a bed forms.  The fluid velocity is correspondingly reduced to near-zero values in the bed, but the net flow of matter is still in the negative $x$-direction.  Upon increasing the gravity effect further, complete flow reversal happens (panel (c)).  These results also suggest the possibility of a near-stationary particle bed for suitable values of both $\Phi$ and $\mygrav$.  A sample of such a result is shown in Figure~\ref{fig:sample_figs1}.
%

\begin{figure}[H]
	\centering
		\subfigure[$\,\,Q_{\mathrm{mixture}}$]{\includegraphics[width=0.49\textwidth]{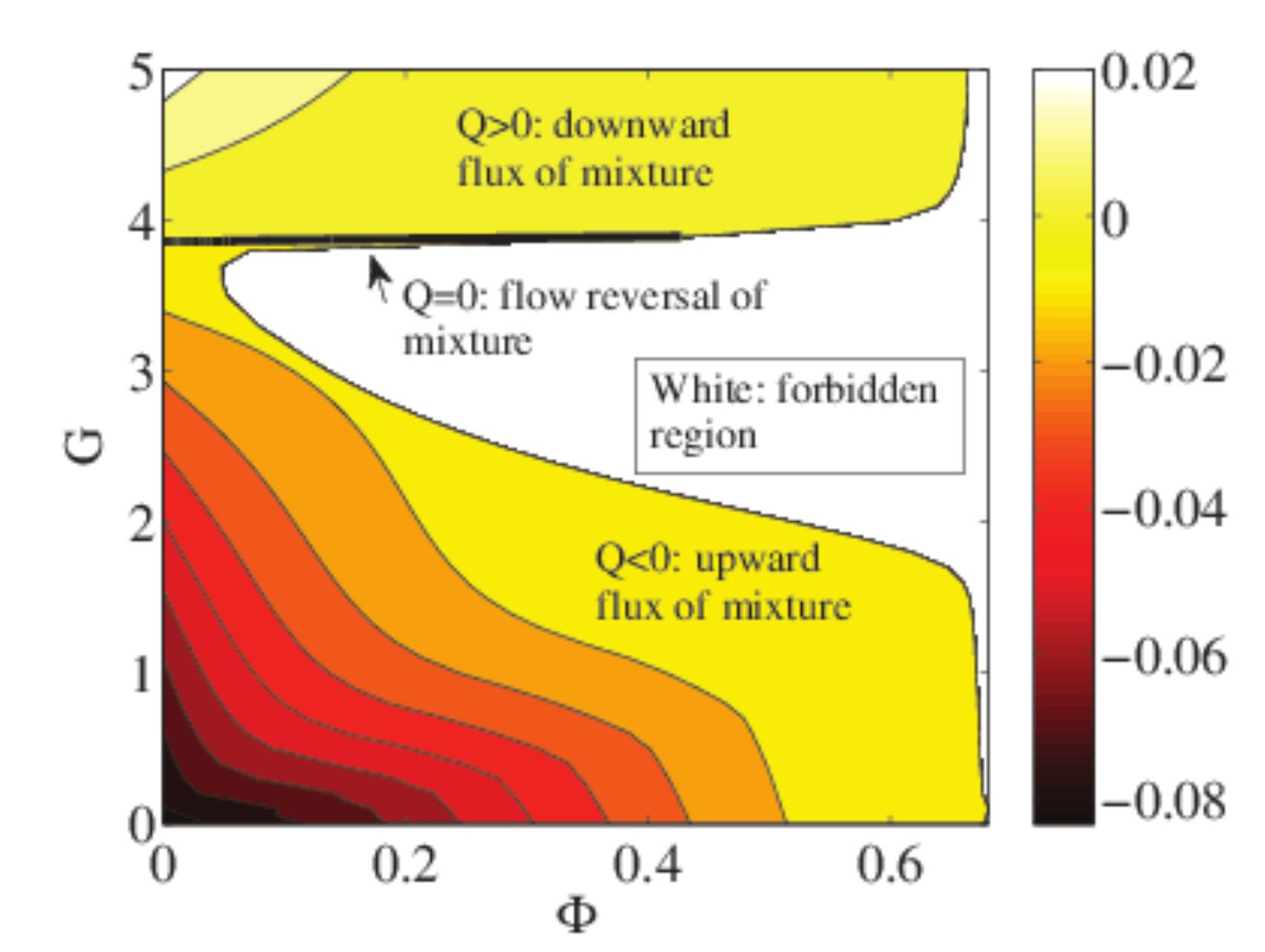}}
		\subfigure[$\,\,Q_{\mathrm{particles}}$]{\includegraphics[width=0.49\textwidth]{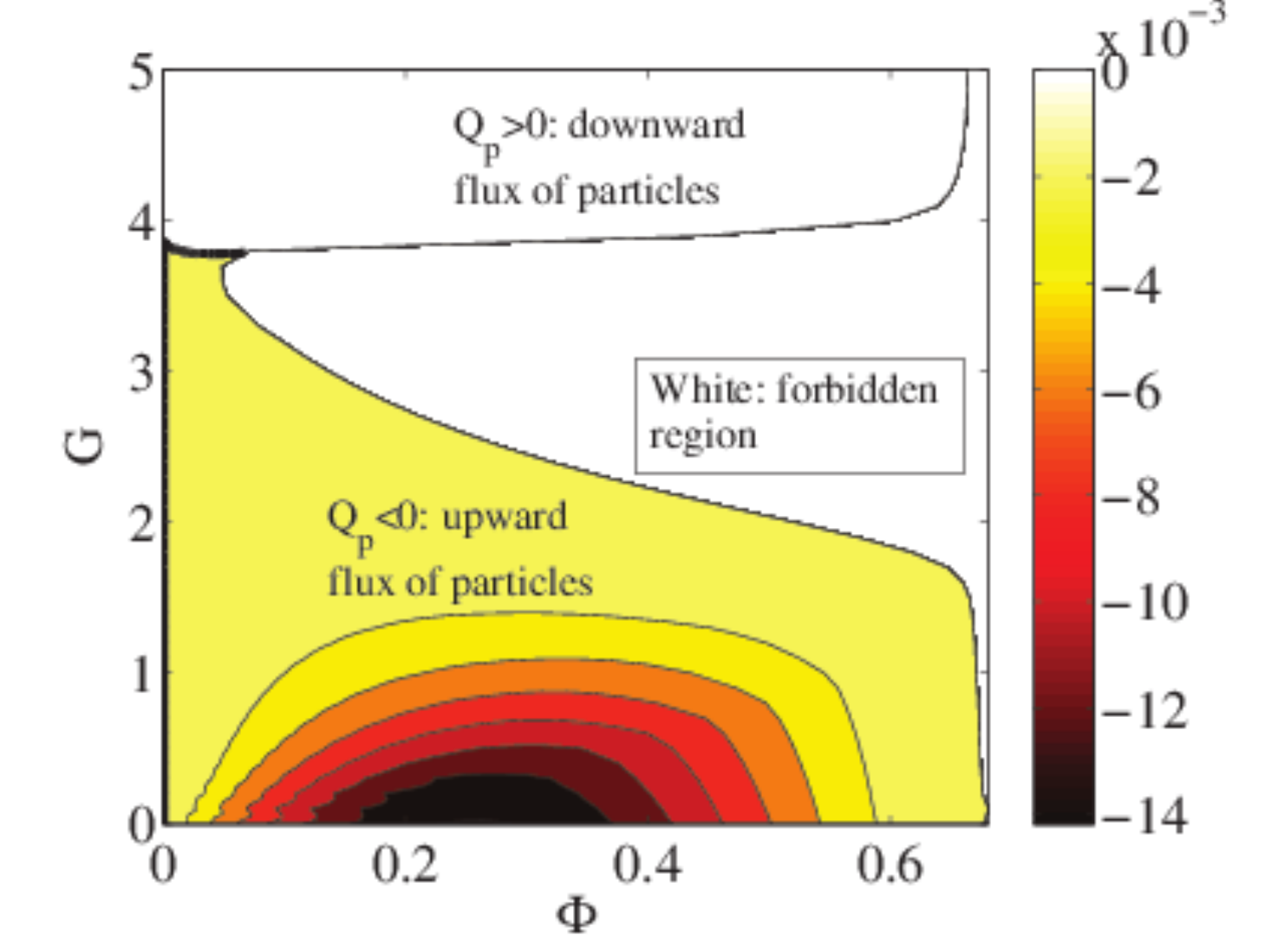}}
		\caption{(a) Mixture flowrate as a function of $(\Phi,\mygrav)$ showing flow reversal line $Q_{\mathrm{mixture}}=0$ at large values of $\mygrav$; (b) Particle flowrate, showing 
		reversal of particle flux at $Q_{\mathrm{particles}}=0$.}
	\label{fig:contours1}
\end{figure}

To understand these trends in a more systematic way, contour plots of the mixture volumetric flow rate and the particle volumetric flow rate were obtained.  These quantities are given by the respective equations
\begin{equation}
Q_{\mathrm{mixture}}=\int_0^1 U(z)\,\mathd z,\qquad Q_{\mathrm{particles}}=\int_0^1 \phi(z)U(z)\,\mathd z;
\end{equation}
the fluid volumetric flow rate is obtained as $Q_{\mathrm{fluid}}=Q_{\mathrm{mixture}}-Q_{\mathrm{particles}}$.  
For simplicity, here and hereafter we will adopt the notation $Q$ and $Q_p$ to denote $Q_{\mathrm{mixture}}$ and $Q_{\mathrm{particles}}$, respectively. 
The corresponding plots are shown in Figure~\ref{fig:contours1}.
In both panels, there is a blank region in parameter space corresponding to large particle volume fractions $\Phi$ and intermediate values of $\mygrav$.  In this region, no solution to the ODE system~\eqref{eq:model_ode} exists.  Physically, this region would correspond to a very high density of particles in a slow-moving or even stationary bed which is unsustainable as an equilibrium solution, and would correspond to a `clogging scenario', wherein the particles overwhelm the flow and lead to a breakdown in the fully-developed flow.  

\subsection{Countercurrent flow and full parameter study}
\label{sec:flow_pattern_map}

\begin{figure}
	\centering
		\includegraphics[width=0.6\textwidth]{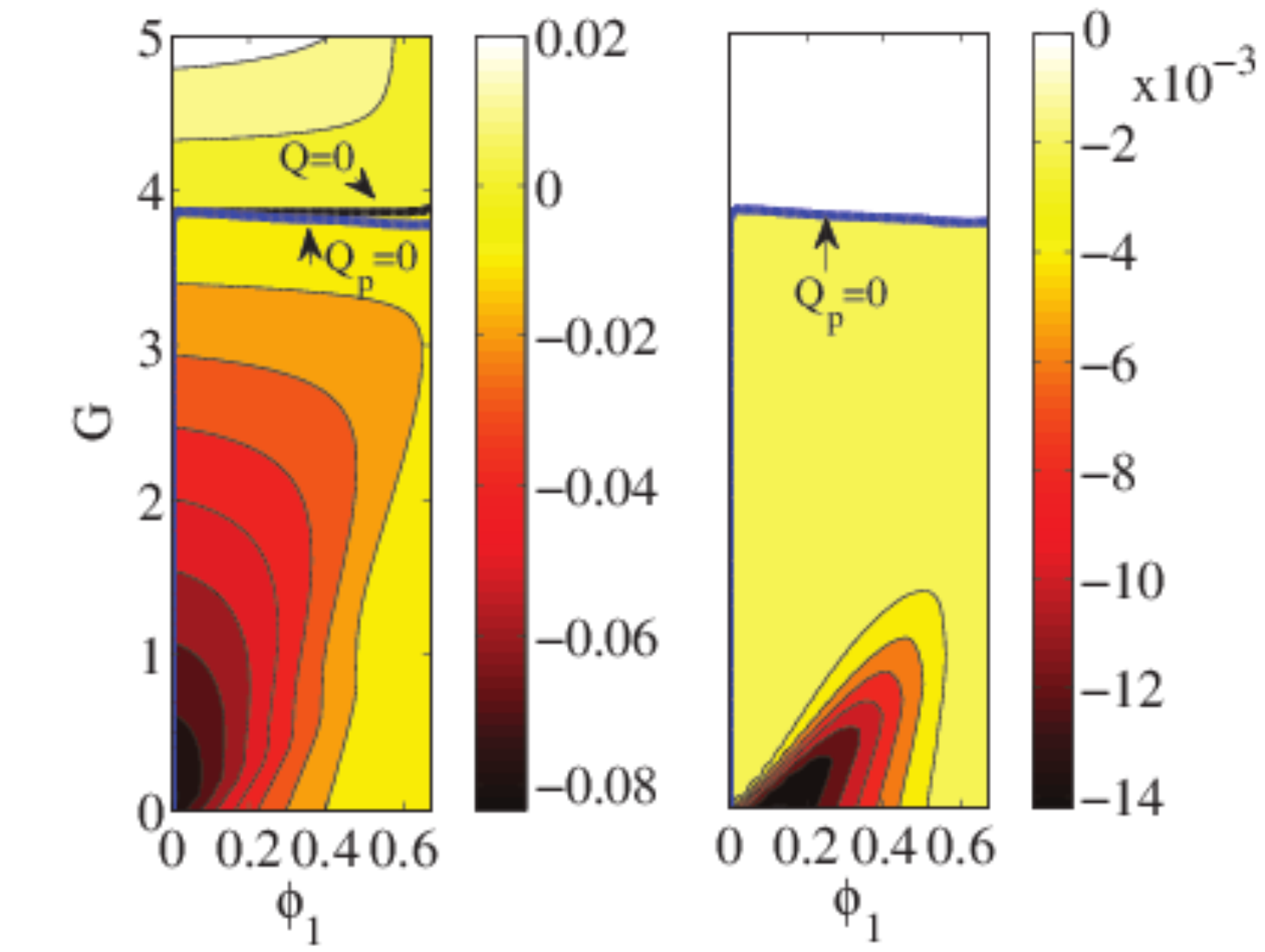}
	\caption{The same as Figure~\ref{fig:contours1}, except in a $(\phi_1,\mygrav)$ parameter space, and showing the non-coincidence of the curves $Q=0$ and $Q_p=0$.  The panel on the left shows $Q$ and the panel on the right shows $Q_p$.  The region marked `DOWN' in the left-hand panel shows the parameter regime for which $Q>0$, corresponding to the mixture moving down the channel.  The region marked `UP' in the same panel shows a parameter regime in which the mixture moves up the channel.  The region enclosed by the two curves corresponds to a scenario wherein the net movement of the mixture is upwards but the particle flux is down, shown below to correspond to countercurrent flow.}
		\label{fig:countercurrent1}
\end{figure}

Referring again to Figure~\ref{fig:contours1}(a),  there is a critical curve $(\Phi_\mathrm{c},{\mygrav}_\mathrm{c})$ (marked in a solid  black line) corresponding to parameter values for which $Q=0$ (panel (a)).  Separately, in panel (b) there is a critical curve which $Q_p=0$.  The curve $Q=0$ corresponds to the situation of complete flow reversal, such that the mixture is transported down the length of the channel (as opposed to direction implied in Figure~\ref{fig:sketch_channel}).  In contrast, the curve $Q_p=0$ corresponds to a reversal in the direction of the particle flux.  These two scenarios are connected but they are not the same.  In particular, the curves $Q=0$ and $Q_p=0$ do not coincide -- see Figure~\ref{fig:countercurrent1}.  For comparison with the parameter studies that follow, the variables in Figure~\ref{fig:countercurrent1} are chosen to be ($\phi_1,\mygrav$), as opposed to $(\Phi,\mygrav)$.

\begin{figure}
	\centering
		\includegraphics[width=0.5\textwidth]{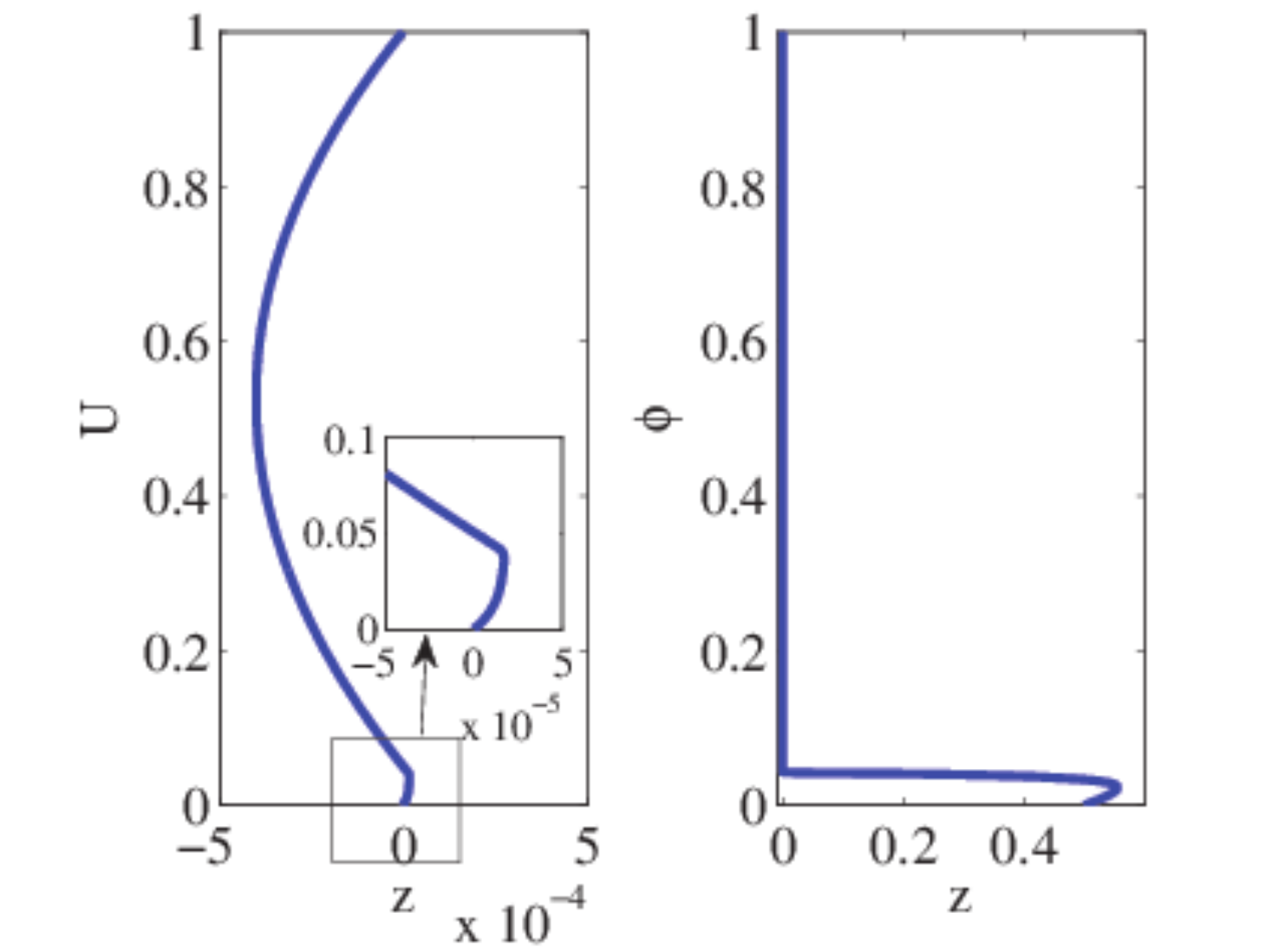}
		\caption{Countercurrent flow regime corresponding to the parameter values $(\phi_1,\mygrav)=(0.5,3.85)$.  Additional parameter values $r=2$, $\alpha=\pi/12$.  The corresponding value of the bulk volume fraction is $\Phi=0.021$.}
	\label{fig:my_u_phi_countercurrent}
\end{figure}

To examine the flow structure in the region enclosed by the curves $Q=0$ and $Q_p=0$ the parameters $(\phi_1,\mygrav)=(0.5,3.85)$ were chosen and the corresponding velocity and volume fraction profiles 
were generated (Figure~\ref{fig:my_u_phi_countercurrent}).
This figure demonstrates that the region enclosed by the curves $Q=0$ and $Q_p=0$ corresponds to a countercurrent flow scenario, where a clear liquid layer is transported upwards and a dense particle-laden layer travels downwards.  Based on the parameter study in Figure~\ref{fig:countercurrent1} it can be expected that complete flow reversal is preceded by a small zone of countercurrent flow (i.e. small in the $(\phi_1,\mygrav)$ parameter space).  The importance of the countercurrent flow to the operation of the system in the steady state can be determined from the area enclosed by the curves $Q=0$ and $Q_p=0$ in the $(\phi_1,\mygrav)$ parameter space, and by the thickness of the corresponding countercurrent layer (e.g. Figure~\ref{fig:my_u_phi_countercurrent}).

\begin{figure}
	\centering
		\subfigure[$\,\,r=2,\alpha=\pi/12$]{\includegraphics[width=0.48\textwidth]{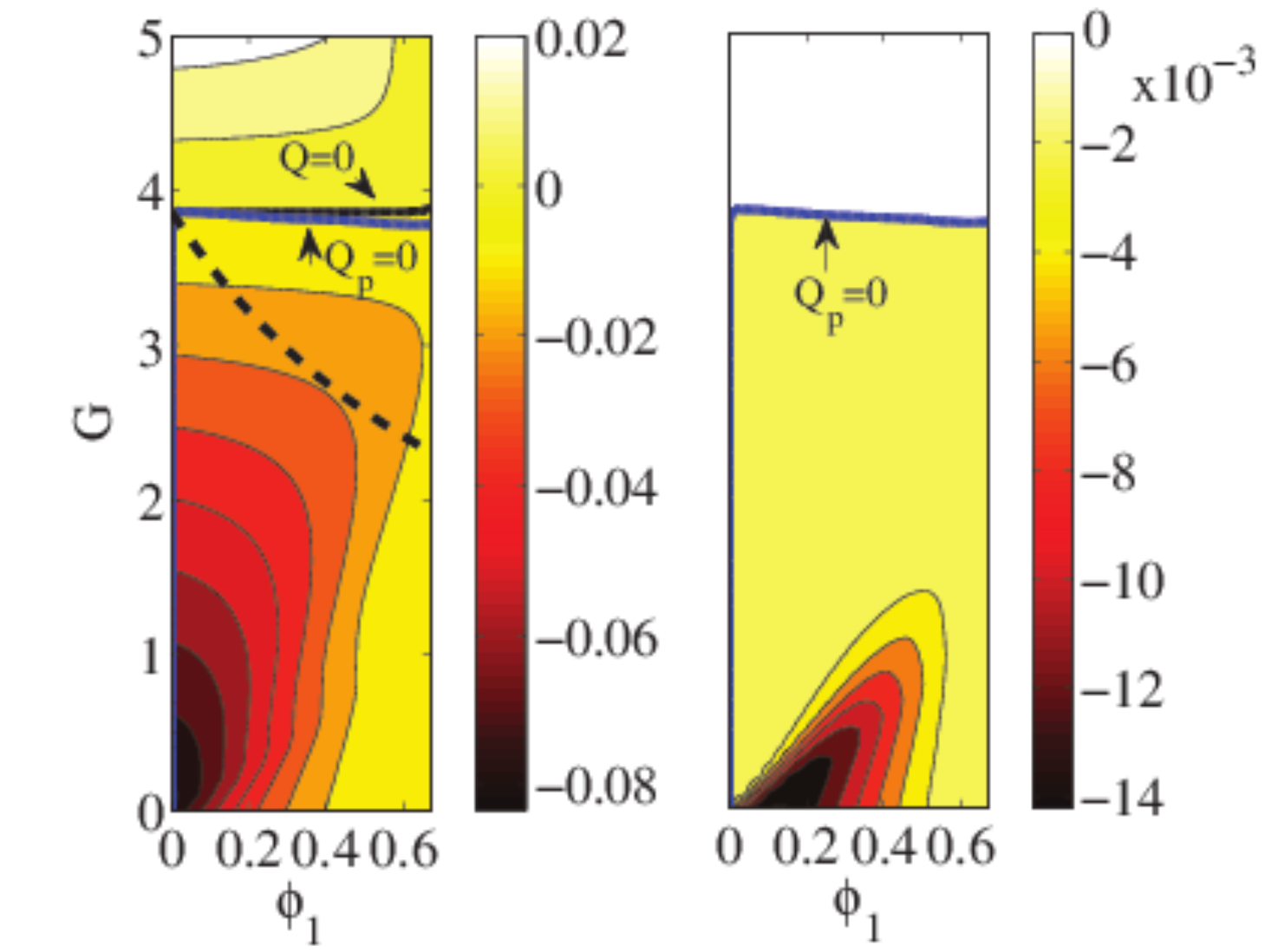}}
		\subfigure[$\,\,r=5,\alpha=\pi/12$]{\includegraphics[width=0.48\textwidth]{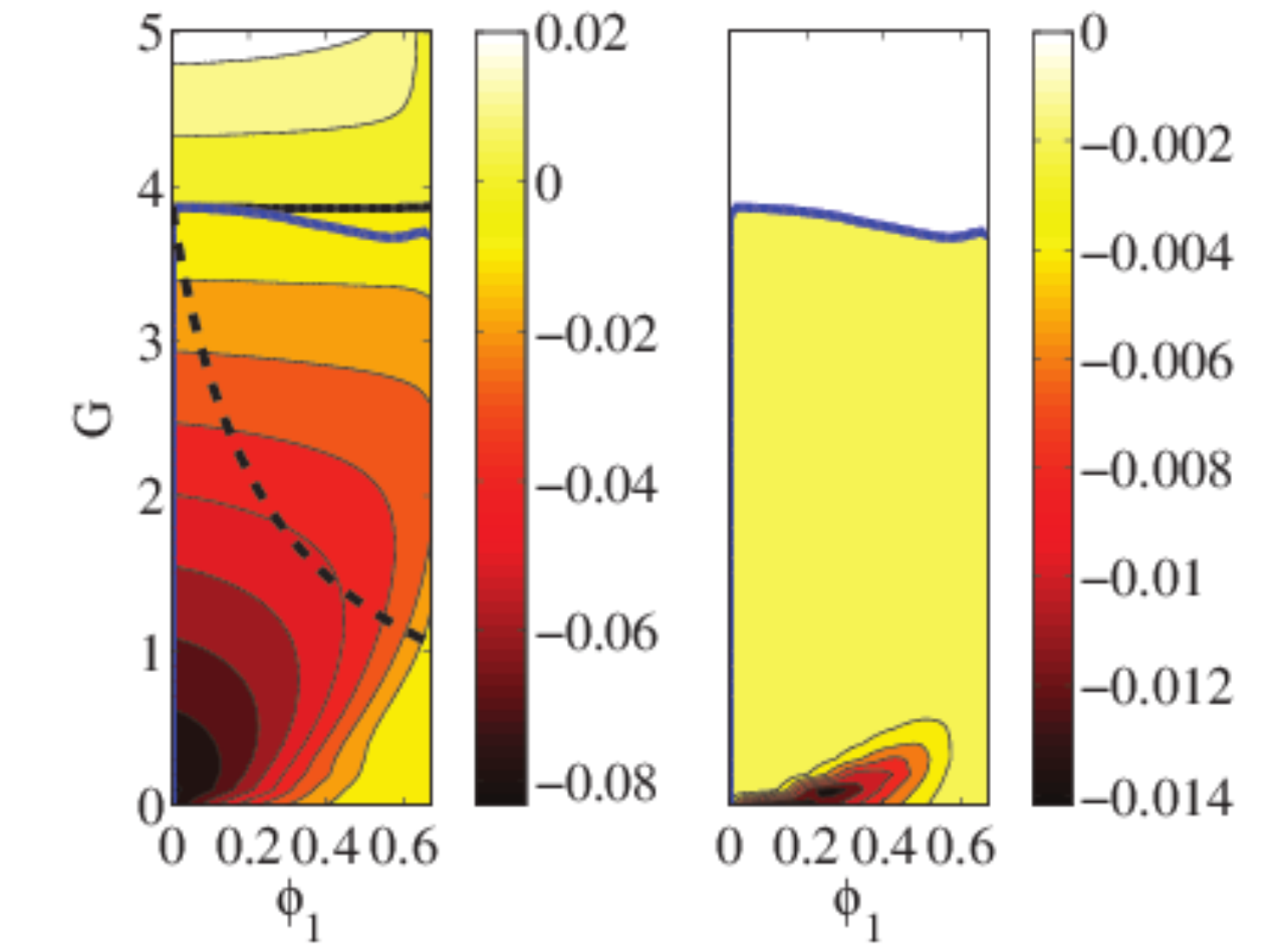}}
		\subfigure[$\,\,r=2,\alpha=\pi/4$]{\includegraphics[width=0.48\textwidth]{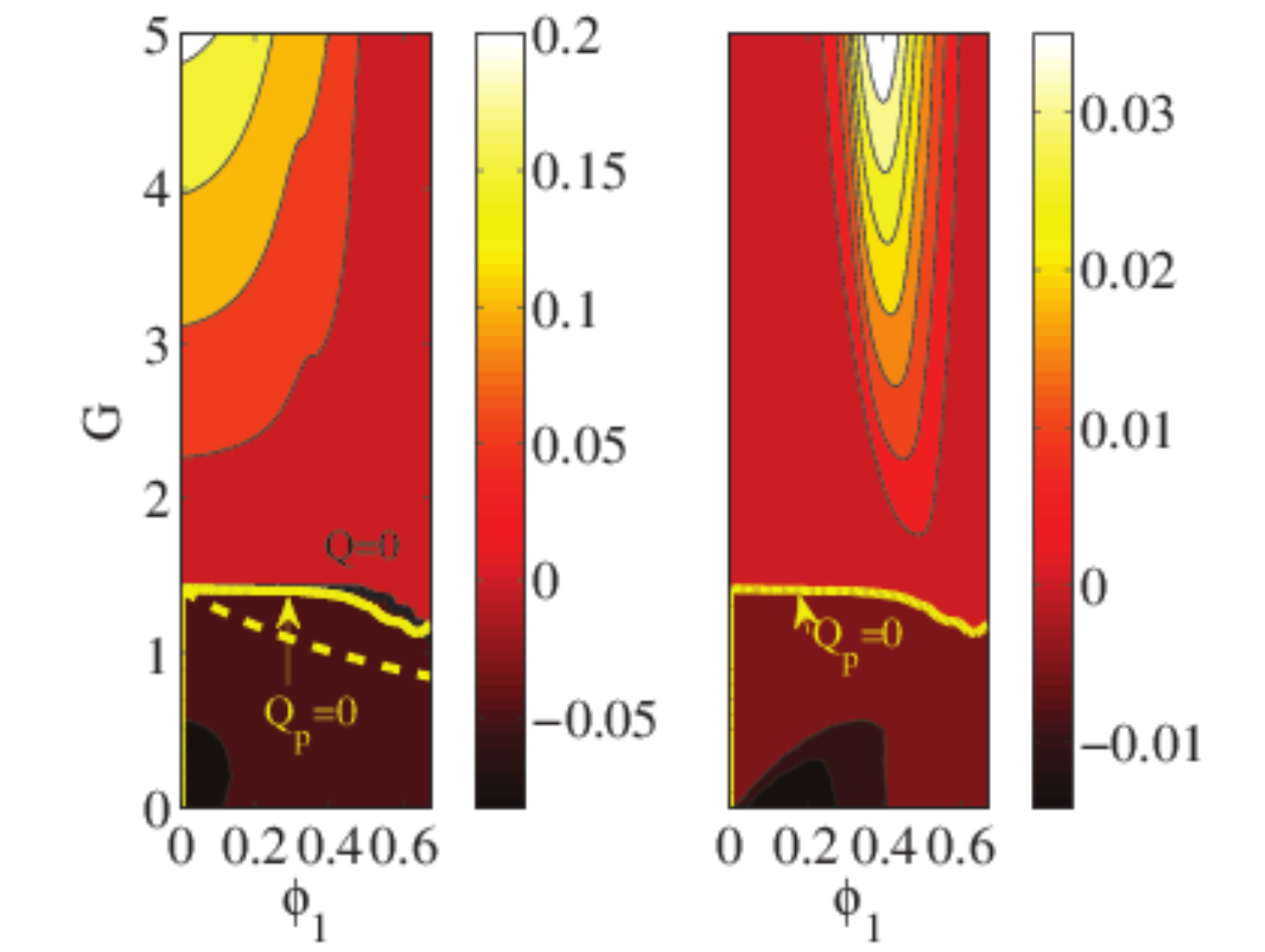}}
		\subfigure[$\,\,r=5,\alpha=\pi/4$]{\includegraphics[width=0.48\textwidth]{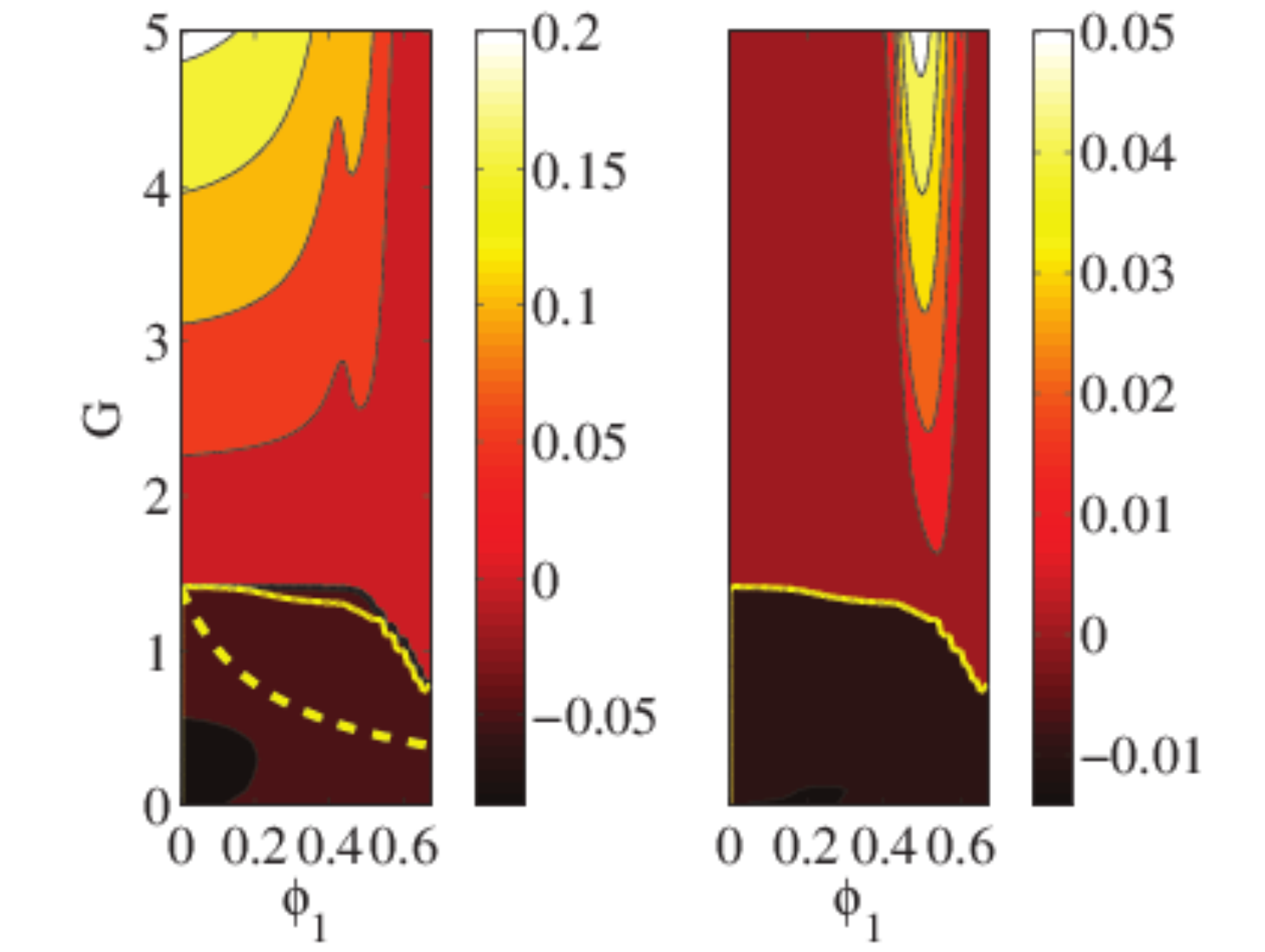}}
		\caption{Flow-pattern maps for various values of $r$ and $\alpha$.  Subfigure (a) shows the previously-considered base case $r=2$, $\alpha=\pi/4$.  Each subfigure contains two panels showing $Q$ on the left and $Q_p$ on the right.  The curves $Q=0$ and $Q_p=0$ and their non-coincidence are shown in each subfigure.  The region enclosed by these curves corresponds to countercurrent flow.
The curve $Q=0$ corresponds to the onset of complete flow reversal; the dashed line shows the lower bound for the onset of complete flow reversal given by Equation~\eqref{eq:gravc}. }
	\label{fig:flow_maps_all}
\end{figure}
\begin{figure}
	\centering
		\subfigure[$\,\,\alpha=\pi/4,r=2$]{\includegraphics[width=0.45\textwidth]{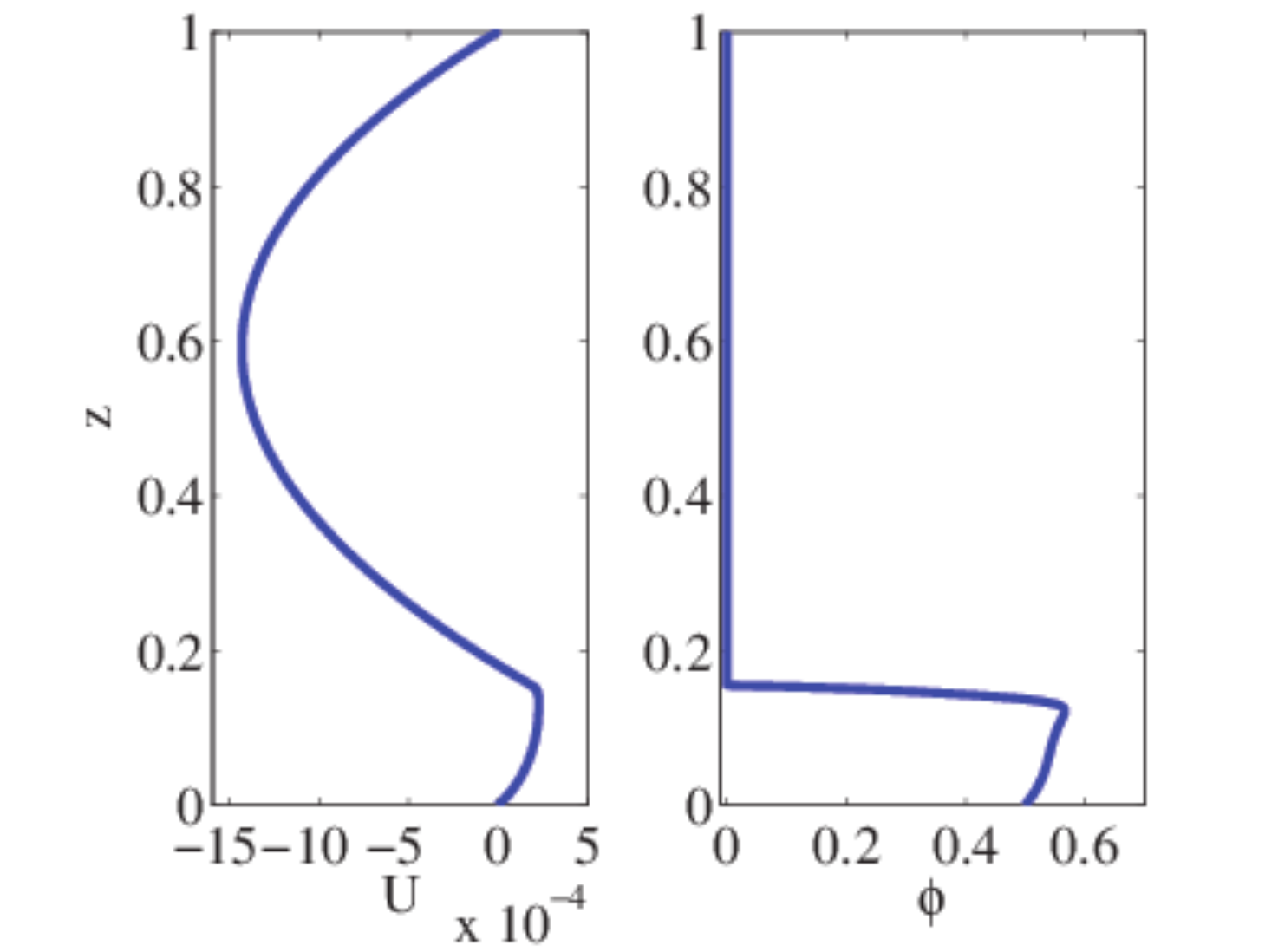}}
		\subfigure[$\,\,\alpha=\pi/4,r=5$]{\includegraphics[width=0.45\textwidth]{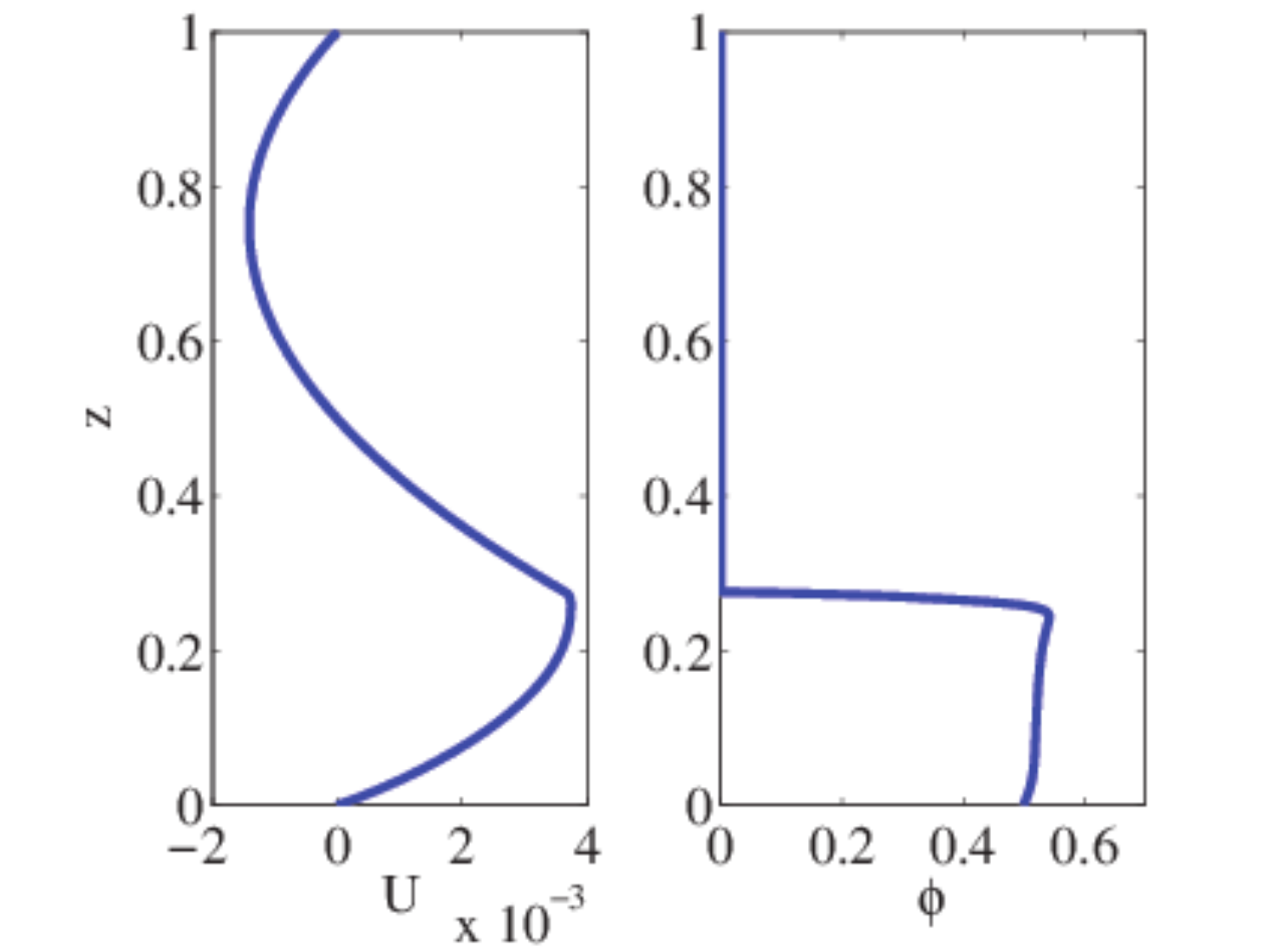}}
		\caption{Countercurrent flow regimes with $\phi_1=0.5$, and $\alpha=\pi/4$, at different density ratios.  (a) $r=2$, $\mygrav=1.39$; (b) $r=5$, $\mygrav=1.35$.}
	\label{fig:my_u_phi_countercurrent1}
\end{figure}

In the above example, it can be seen that the countercurrent region is not very large in parameter space, nor yet very important to the actual flow structure.  Therefore, we complete a further parameter study wherein we investigate whether the regime of countercurrent flow can be extended by varying other flow parameters, in particular the inclination angle $\alpha$ and the particle density $r$.
The resulting mixture flowrate and particle flux for various values of $(\alpha,r)$ are shown in the flow-pattern maps in Figure~\ref{fig:flow_maps_all}.
Changes to the density ratio and the angle of inclination with respect to the base case $(\alpha,r)=(\pi/12,2)$ produce visible effects in the structure of the flow-pattern maps.   In particular:
\begin{itemize}
\item An increase in the angle of inclination $\alpha$ means that the critical value of $\mygrav$ for the onset of complete flow reversal is lowered with respect to the base case: from $\mygrav\approx 4$ to $\mygrav \approx 1.5$.   This can be seen by comparing Figures~\ref{fig:my_u_phi_countercurrent1}(a) and (c).
\item An increase in $\alpha$ causes a downward surge in the particle flux at intermediate volume fractions.  
This can be seen by comparing Figures~\ref{fig:my_u_phi_countercurrent1}(a) and (c) again: for $\alpha=\pi/12$ the quantity $Q_p$ attains large negative values at $\phi_1\approx 0.2$ and $\mygrav=0$, corresponding to an upward surge in the particle flux.  In contrast, for $\alpha=\pi/4$, $Q_p$ attains large positive values for large values of $\mygrav$ and $\phi_1\approx 0.5$ and corresponds to a downward surge in the particle flux.  
\item An increase in the density ratio $r$ means that the region in parameter space in which countercurrent flow is observed increases - albeit only slightly.
This can be seen by comparing Figures~\ref{fig:my_u_phi_countercurrent1}(a) and (b) and separately, Figures~\ref{fig:my_u_phi_countercurrent1}(c) and (d).
\item Increases in either the angle of inclination or the density ratio lead to a more prominent countercurrent flow structure, as evidenced by the contrast between 
Figure~\ref{fig:my_u_phi_countercurrent}  and
Figure~\ref{fig:my_u_phi_countercurrent1}.
\end{itemize}
We have verified that increasing the angle of inclination beyond $\alpha=\pi/4$ does not introduce any further qualitative changes to the flow-pattern map.
%

%

\begin{figure}[H]
	\centering
		\subfigure[]{\includegraphics[width=0.45\textwidth]{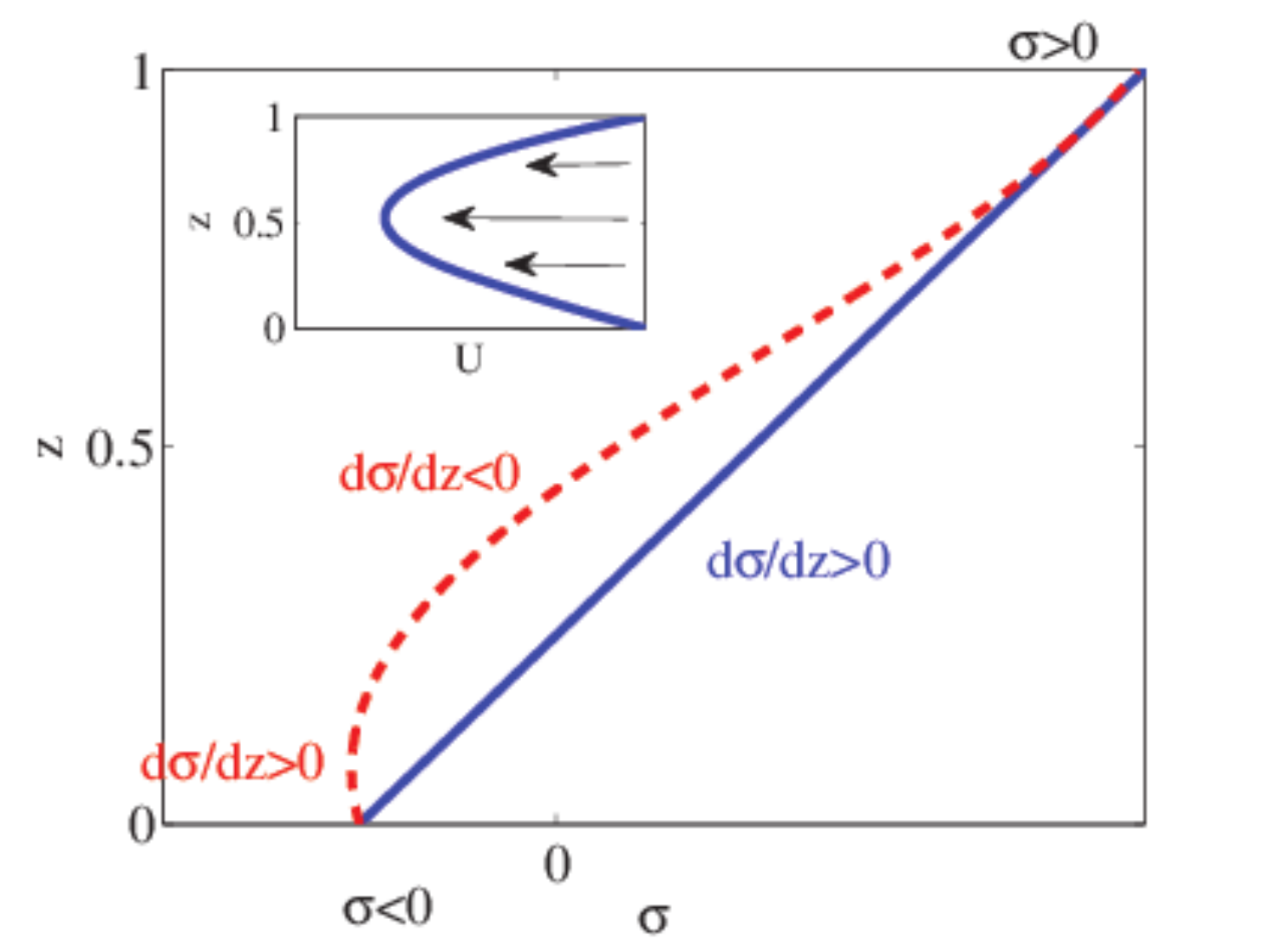}}
		\subfigure[]{\includegraphics[width=0.45\textwidth]{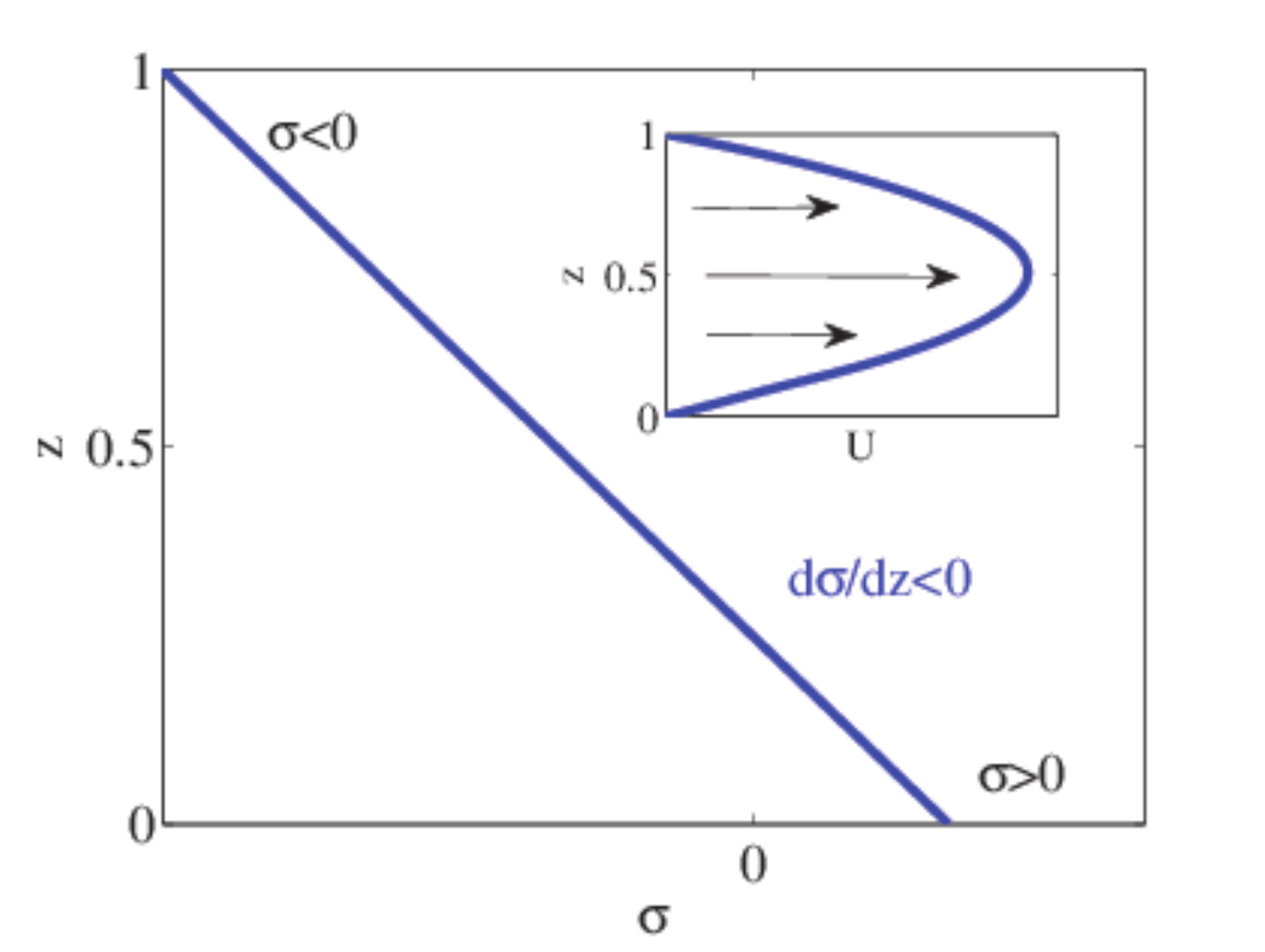}}
		\caption{Conditions for flow reversal: $\sigma(z)$ switches from increasing to decreasing at a threshold value of $\mygrav$ given by Equation~\eqref{eq:gravc}.  The situation in panel (a) (broken line) then pertains.  Only when $\mygrav$ is increased beyond the threshold value of $\mygrav$ does $\sigma(z)$ change to being an strictly decreasing function whereupon complete flow reversal occurs.}
	\label{fig:thm_sketch}
\end{figure}

Finally, it is possible to obtain a lower bound on the critical value $\mygrav_\mathrm{c}$ necessary for the onset of complete flow reversal.  The result is that starting with $\mygrav=0$, and for a fixed value of $\phi_1$, it is necessary to increase $\mygrav$ at least to the value
\begin{equation}
\mygrav=\frac{1}{\sin\alpha}\frac{1}{r\phi_1+(1-\phi_1)}
\label{eq:gravc}
\end{equation}
in order for the flow to reverse completely upwards to downwards, that is, the true critical value $\mygrav_\mathrm{c}$ for the onset of flow reversal is bounded below such that
\[
\mygrav_\mathrm{c}\geq \frac{1}{\sin\alpha}\frac{1}{r\phi_1+(1-\phi_1)}.
\]
One can see this as follows: for a scenario involving a strong upward flow, we will have $U(z)<0$, with $\sigma$ a monotone-increasing function (Figure~\ref{fig:thm_sketch}(a), solid line).  In contrast, for a strong downward flow the opposite situation will pertain (Figure~\ref{fig:thm_sketch}(b)).  There is a crossover point where $(\mathd\sigma/\mathd z)_0=0$ -- this condition gives Equation~\eqref{eq:gravc}.  However, complete flow reversal does not happen exactly at this point, since the scenario shown in Figure~\ref{fig:thm_sketch}(a) (dashed line) may pertain.  Therefore, $\mygrav$ must be at or beyond the point given in Equation~\eqref{eq:gravc} in order for the complete flow reversal to occur.

We have checked that the critical curve for the onset of complete flow reversal lies above the lower bound in Equation~\eqref{eq:gravc} for each of the parameter cases considered in Figure~\ref{fig:flow_maps_all} -- the bound is shown as a dashed curve in that figure.  The bound is by no means sharp, although it sharpens  with increasing inclination angle; the sharpness of the bound also increases with decreasing density ratio.

\section{Discussion and conclusions}
\label{sec:disc}

So far we have considered only the case wherein both the suspending fluid and the mixture possess a Newtonian rheology.  In practice, these are unrealistic assumptions -- particularly the latter.  However, since the rheological properties of many fluids can be fitted to a Herschel--Bulkley model, the applicability of the regularized diffusive-flux equations is recovered by introducing the Herschel--Bulkley model into the momentum-balance equation.
We outline here the main changes required to implement this step.  Crucially, constitutive modelling of four quantities is required to close the non-Newtonian version of Equation~\eqref{eq:model_ode}.  These are the mobility function, the consistency, power-law index for the rheological model, and the Bingham number, detailed here as follows:
\begin{itemize}
\item  A model for the settling viscosity in a Herschel--Bulkley fluid.  Thus, equation~\eqref{eq:j_grav} must be replaced by 
\begin{equation}
\bm{J}_\mathrm{g}=-\phi M_{NN}(\phi,z) \rFr^{-2}(\sin\alpha,0,-\cos\alpha),
\end{equation}
where the mobility function $M_{NN}(\phi,z)$ is to be determined.
\item The Herschel--Bulkley model for the suspension, given by a (non-dimensional) constitutive relation of the  form
\begin{equation}
\mu(\phi)=k(\phi)\dot\gamma^{n(\phi)-1}+\frac{\rBn(\phi)}{\dot\gamma}.
\end{equation}
Here, $k(\phi)$ is the non-dimensional consistency, $n(\phi)$ is the power and $\rBn(\phi)$ is the non-dimensional Bingham number.  Thus, $(k(0),n(0),\rBn(0))$ correspond to a reversion to the rheological properties of the pure suspending fluid, i.e. $\phi=0$.   These considerations lead to a constitutive relation for the viscous component of the suspension stress tensor, $\vecsig=\mu(\phi)\dot{\bm{\gamma}}$. 
\end{itemize}
An example of how the rheological parameters $(n,\rBn,k)$ can be measured as a function of the volume fraction and then fitted to a Herschel--Bulkley model can be found in Reference~\cite{mueller2009rheology}.
Once these relations are supplied, the ODE system~\eqref{eq:model_ode} can be reformulated as follows:
\begin{subequations}
\begin{eqnarray}
\frac{\mathd U}{\mathd z}&=&\begin{cases}\mathrm{sign}(\sigma)\left(\frac{|\sigma|-\rBn(\phi)}{k(\phi)}\right)^{1/n(\phi)},& |\sigma|>\rBn(\phi),\\
                                        0,&\text{otherwise},
																				\end{cases}\\
\frac{\mathd\sigma}{\mathd z}&=&1-\rRe \, \rFr^{-2}\left[r\phi+(1-\phi)\right]\sin\alpha,\\
\frac{\mathd\phi}{\mathd z}&=&\begin{cases}0,&\text{if }\phi=0,\text{ or }\phi=\phi_\mathrm{m},\\
\frac{-\phi \frac{\sigma}{\hat\sigma}\frac{\mathd\sigma}{\mathd z}-\tfrac{1}{\dc}\rFr^{-2} M_{NN}(\phi,z)\cos\alpha}{\hat\sigma\left[1+2\left(\frac{\dv-\dc}{\dc}\right)\frac{\phi}{\phi_\mathrm{m}-\phi}-\frac{\epsilon^2}{\hat\sigma^2}\frac{\mathd\sigma}{\mathd z} \phi \rRe \, \rFr^{-2}(r-1)\sin\alpha\right]},&\text{otherwise}.\end{cases}
\end{eqnarray}%
\label{eq:model_ode_nn}%
\end{subequations}%
A thorough analysis of these equations is left for future work, although we note in passing that the parameter space of Equations~\eqref{eq:model_ode_nn} is much enlarged with respect to the present study.

A second future aspect concerns the study of the suspension under transient simulations, via direct numerical simulation.  It will be straightforward formally to couple the regularized momentum-balance equation~\eqref{eq:hydro} to the diffusive-flux transport equation~\eqref{eq:flux}, although a thorough analysis of the resulting system may be required to ensure a robust numerical solution, as the regularization introduces higher-order velocity derivatives into the diffusive-flux transport equation.  
Finally, the study of the system behavior under transient conditions can be linked to complex geometries such as those found in oil-well-drilling by investigating the linear stability of a suspension flow in a eccentric Taylor--Couette geometry.  Such work has already been done for ordinary Newtonian flows~\cite{leclercq2013temporal}; it would be of great theoretical and practical interest to extend this to suspensions such as the model suspension considered herein.

Summarizing the present work, we have introduced a regularized diffusive-flux model for a viscous suspension.  The regularization removes the unphysical cusp in the volume-fraction profile at points where the shear rate vanishes.  It also introduces an explicit dependence on the particle radius into the problem.  The model enables a complete exploration of the parameter space involving the ratio $\mygrav=\rf g/|\mathd P/\mathd L|$, the density ratio $r=\rp/\rf$, the angle of inclination, and the bulk volume fraction.  The flow regimes are mapped as a function of these parameters and conditions for complete flow reversal and countercurrent flow are identified.

\appendix

\section{Root-mean-square average of the shear stress over a particle}
\label{sec:app:derivation}

We demonstrate here how Equation~\eqref{eq:sig_avgxxx} in the main text amounts to the root-mean-square average shear stress over the particle.  For, consider a particle located at $\vecx_0$ in the domain.  The mixture shear stress at this point is $\sigma(\vecx_0)$.  In the present context of unidirectional flow,  $\sigma=\pm\mu(\phi)\dot\gamma$ is the scalar-valued signed shear stress of the mixture, and $\dot\gamma=|\mathd U/\mathd z|$ is the rate of strain.  We emphasize that the approach in this appendix can be generalized to arbitrary flows.
Instead of using $|\sigma|$ in the diffusive-flux model (e.g. as in Equation~\ref{eq:dphidz_dodgy}), we consider instead the following averaged shear stress, where the average is taken in the $L^2$ norm:
\begin{equation}
\hat\sigma(\vecx_0)=\left[\frac{1}{|S(\vecx_0,a)|}\int_{S(\vecx_0,a)}\sigma^2(\vecx)\mathd S\right]^{1/2}
\label{eq:app:sig_def}
\end{equation}
where $S(\vecx_0,a)$ is the sphere of centre $\vecx_0$ and radius $a$ and $|S(\vecx_0,a)|$ is the surface area of the sphere (i.e. particle surface area, equal to $4\pi a^2$ in three dimensions).   Also, $\mathd S$ denotes an infinitesimal patch of area on the boundary sphere $S(\vecx_0,a)$.  In what follows, it is important to compute the averages in a fully three-dimensional manner: although the assumed flow is unidirectional, the particles are embedded in a three-dimensional domain, and hence, the three-dimensional averaging technique is necessary.

Because the particles are assumed to have a small radius in comparison to the channel height, the shear stress is expanded in a Taylor expansion, centred at the particle centre, to second order:
\begin{multline}
\sigma(\vecx)=\sigma(\vecx_0)+(x_i-x_{0i})b_i+\tfrac{1}{2}(x_i-x_{0i})(x_j-x_{0j})A_{ij},\\
b_i=\frac{\partial\sigma}{\partial x_i}\bigg|_{\vecx_0},\qquad A_{ij}=\frac{\partial^2\sigma}{\partial x_i\partial x_j}\bigg|_{\vecx_0}
\end{multline}
where we sum over repeated indices.
We denote the radicand in Equation~\eqref{eq:app:sig_def} by $\mathcal{I}$; we have
\begin{eqnarray*}
\mathcal{I}&=&\frac{1}{|S(\vecx_0,a)| }\int_{S(\vecx_0,a)}\sigma^2(\vecx)\mathd S,\\
&=&\frac{1}{|S(\vecx_0,a)|}\int_{S(\vecx_0,a)} \left[\sigma(\vecx_0)+(x-x_{0i})b_i+\tfrac{1}{2}(x-x_{0i})(x-x_{0j}) A_{ij}\right]^2 \mathd S,\\
&=&[\sigma(\vecx_0)]^2+2|S(\vecx_0,a)|^{-1}\sigma(\vecx_0)b_i\int_{S(\vecx_0,a)}(x_i-x_{0i})\mathd S\\
&\phantom{=}&\phantom{aaa}+|S(\vecx_0,a)|^{-1}\left[b_i\int_{S(\vecx_0,a)}(x_i-x_{0i})\mathd S\right]^2\\
&\phantom{=}&\phantom{aaaaa}+|S(\vecx_0,a)|^{-1}\sigma(\vecx_0)A_{ij}\int_{S(\vecx_0,a)}(x-x_{0i})(x-x_{0j})\mathd S+\text{higher-order terms}.
\end{eqnarray*}
Doing the integrals and neglecting the higher-order terms, this works out to be 
\begin{eqnarray}
\mathcal{I}&=&\sigma(\vecx_0)^2+\tfrac{1}{3}a^2 |\bm{b}|^2+\tfrac{1}{3}a^2\sigma(\vecx_0)A_{ij}\delta_{ij},\nonumber\\
&=&\sigma(\vecx_0)^2+\tfrac{1}{3}a^2 [\nabla\sigma(\vecx_0)]^2+\tfrac{1}{3}a^2\sigma(\vecx_0)\nabla^2\sigma(\vecx_0).
\end{eqnarray}
valid to second order in a Taylor expansion.  For a unidirectional flow, this reduces to
\begin{equation}
\mathcal{I}=\left[\sigma(\vecx_0)\right]^2+\tfrac{1}{3}a^2\left[\sigma'(\vecx_0)\right]^2+
\tfrac{1}{3}\sigma(\vecx_0)a^2\left[\sigma''(\vecx_0)\right]^2,
\label{eq:app:problem}
\end{equation}
where $\sigma'=\mathd\sigma/\mathd z$ etc.

The second-order derivative is problematic in Equation~\eqref{eq:app:problem}.  However, it can safely be ignored.  For, the terms proportional to $a^2$ are important only when $\sigma(\vecx_0)\rightarrow 0$.  In this limit, the term involving the second-order derivative tends to zero as well.  In other words, we have the following approximation:
\[
\mathcal{I}\approx\begin{cases} \tfrac{1}{3}a^2\left[\sigma'(\vecx_0)\right]^2\text{ for }\sigma(\vecx_0)\rightarrow 0,\\
[\sigma(\vecx_0)]^2\text{ otherwise},\end{cases}
\]
meaning that the approximation
\[
\mathcal{I}\approx \left[\sigma(\vecx_0)\right]^2+\tfrac{1}{3}a^2\left[\sigma'(\vecx_0)\right]^2
\]
is uniformly valid, hence
\[
\hat\sigma(\vecx_0)\approx \sqrt{ \left[\sigma(\vecx_0)\right]^2+\tfrac{1}{3}a^2\left[\sigma'(\vecx_0)\right]^2}
\]
i.e. Equation~\eqref{eq:sig_avgxxx} in the main text is recovered.  

\subsection*{Acknowledgements}

This work arose from the  102$^{\text{nd}}$ European Study Group with Industry, hosted by University College Dublin, Ireland in July 2014.  The work is based on the project provided by the International Research Institute Stravanger (IRIS), Norway.  The authors acknowledge the presentation of the problem by Fionn Iversson and Johnny Petersen.  The authors also acknowledge the participation of the members of the study group, including Panagiotis Giounanlis, Susana Gomes, Dan Lucas, Orlaith Mannion, Rachel Mulungye, and Brendan Murray. R.B. acknowledges the support of  
 Science Foundation Ireland under grant 12/IA/1683 and the support of the Irish Reserach Council under the `New Foundations' scheme (2014).

\subsection*{References}

\end{document}